\documentclass[aps,prd,nofootinbib,preprintnumbers,twocolumn,superscriptaddress,floatfix]{revtex4-2}
\usepackage{amsmath,amssymb}
\usepackage{graphicx}
\usepackage[dvipsnames]{xcolor}
\usepackage{mathtools}
\usepackage{physics}
\usepackage[utf8]{inputenc}
\usepackage{tensor}
\usepackage{comment}
\graphicspath{ {./images/} }
\usepackage{tikz}
\usetikzlibrary{calc}

\usepackage[normalem]{ulem}
\usepackage[setpagesize=false,colorlinks=true,citecolor=TealBlue]{hyperref}
\usepackage{placeins}
\begin{document}

\newcommand{\tcb}{\textcolor{blue}}
\newcommand{\DM}{\text{\tiny DM}}
\newcommand{\REM}{\text{\tiny REM}}
\newcommand{\PI}{\texttt{P1}}
\newcommand{\PII}{\texttt{P2}}
\newcommand{\PIII}{\texttt{P3}}
\newcommand{\obs}{\text{obs}}
\newcommand{\C}{C_{\nu\gamma}}
\newcommand{\pbh}{\text{\tiny PBH}}
\newcommand{\lpbh}{$\ell$PBH}
\renewcommand{\d}{\text{d}}
\newcommand{\maxi}{\text{\tiny{max}}}
\newcommand{\LQG}{\text{\tiny LQG}}
\newcommand{\GW}{\text{\tiny{GW}}}
\newcommand{\tot}{\text{\tiny tot}}
\newcommand{\mi}{{m_{\pbh}^{(\rm i)}}}
\newcommand{\mss}{m_\star}
\newcommand{\Ni}{N_\text{i}}
\newcommand{\Nb}{N_\text{b}}
\newcommand{\Nr}{N_\text{r}}
\newcommand{\Nd}{N_\text{dom}}
\newcommand{\dom}{\text{dom}}
\newcommand{\Neff}{N_\text{eff}}
\newcommand{\CDM}{\text{\tiny CDM}}
\newcommand{\ogwe}{\Omega_\text{\tiny GW}^\text{evap}}
\newcommand{\ogwo}{\Omega_{\text{\tiny GW},0}}
\newcommand{\oro}{\Omega_{\text{rad},0}}
\newcommand{\rgh}{\rho_\gamma^\text{\tiny H}}
\newcommand{\bgw}{\beta_\GW}

\def\ga{\mathrel{\raise.3ex\hbox{$>$\kern-.75em\lower1ex\hbox{$\sim$}}}}
\def\la{\mathrel{\raise.3ex\hbox{$<$\kern-.75em\lower1ex\hbox{$\sim$}}}}

\def\be{\begin{equation}}
\def\ee{\end{equation}}
\def\bea{\begin{eqnarray}}
\def\eea{\end{eqnarray}}

\def\betap{\tilde\beta}
\def\del{\delta_{\rm PBH}^{\rm local}}
\def\Msun{M_\odot}
\def\Rcl{R_{\rm clust}}
\def\fPBH{f_{\rm PBH}}

\newcommand{\Mpl}{M_P} 
\newcommand{\mpl}{m_\mathrm{pl}} 

\preprint{KCL-PH-TH-2025-24}

\title{Signatures of loop quantum gravity in primordial black hole cosmologies}

\author{Antoine Dierckx}

\affiliation{Instituto de Estructura de la Materia, IEM-CSIC, C/ Serrano 121, 28006 Madrid, Spain}
\author{S\'ebastien Clesse}

\affiliation{Service de Physique Th\'eorique, Universit\'e Libre de Bruxelles (ULB), Boulevard du Triomphe, CP225, 1050 Brussels, Belgium.}

\author{Francesca Vidotto}

\affiliation{Instituto de Estructura de la Materia, IEM-CSIC, C/ Serrano 121, 28006 Madrid, Spain}


\begin{abstract}
     The possibility that Dark Matter (DM) is partially or totally constituted by stable Planckian remnants of light Primordial Black Holes (PBHs), suggested for instance by Loop Quantum Gravity (LQG), is investigated.  
     Distinct phenomenological regimes are identified, including scenarios that trigger an early matter-dominated epoch. New constraints are derived on the initial PBH and final remnant abundances.
     We show that a significant initial abundance of PBHs lighter than $10^3$ kg would overproduce Planckian relics, implying that any observational evidence for such PBHs would challenge models with quasi-stable remnants.   
     Conversely, the products of Hawking radiation from PBHs with masses between $10^3$ and $10^{12}$ kg impose that Planckian relics could only be a highly subdominant DM component. 
     We identify a PBH mass around $10^3$ kg for which Hawking evaporation naturally reheats the Universe while the remnants entirely constitute the present-day DM. Such a scenario does not require fine-tuning the initial abundance of PBH of this mass, which could range from $10^{-10}$ to order one. These early-Universe cosmologies yield distinct observational signatures: scalar-induced gravitational waves sourced by primordial or Poisson fluctuations that are amplified by the early PBH-dominated era.  Current and future observations of LIGO/Virgo/KAGRA, the Einstein Telescope and LISA, as well as probes of the effective number of relativistic degrees of freedom, can be used to probe and constrain the initial PBH abundance and the present-day abundance of Planckian relics.\\
\end{abstract}

\maketitle
\section{Introduction}\label{sec:Introduction}

Nearly one century after the discovery of its first observational evidence, the nature of dark matter (DM), accounting for about 85\% of the matter in the Universe, remains a deep mystery.  Even though most search efforts have gone in the direction of the existence of a new particle, the advent of gravitational-wave (GW) astronomy has strongly boosted an alternative class of DM candidates, primordial black holes (PBHs)~\cite{Bird:2016dcv,Clesse:2016vqa,Sasaki:2016jop}.  PBHs may have formed in the early Universe~\cite{Hawking:1971ei,Carr:1974nx,Meszaros:1975ef,Chapline:1975ojl} due to the gravitational collapse of primordial curvature fluctuations, for instance produced during inflation (see~\cite{Byrnes:2025tji,LISACosmologyWorkingGroup:2023njw} for recent reviews including possible formation mechanisms). There exists a variety of constraints on the PBH abundance~\cite{Carr:2020xqk} that coexist with possible observational clues~\cite{Carr:2023tpt,Carr:2019kxo,Clesse:2017bsw}.  

Below a mass of about $10^{12}$ kg, it is usually considered that PBHs should have totally evaporated through Hawking radiation.  Nevertheless, when one goes beyond this semi-classical description, it is plausible that PBHs form stable remnants that may contribute to a sizeable fraction or even the totality of the DM~\cite{MacGibbon:1987my}. For instance, the memory burden effect~\cite{Dvali:2020wft} has recently received a lot of attention and predicts black hole relics of about half their initial mass.  Alternatively, accounting for quantum gravity effects can also lead to Planckian relics~\cite{Bianchi:2018mml}, which 
might be directly detected~\cite{Christodoulou:2023hyt}  
and that also constitute a good DM candidate.  This is expected in Loop Quantum Gravity (LQG)~\cite{Rovelli:2024sjl}, where the black hole area becomes quantized.  The PBH evaporation then leads to Planckian remnants that are quantum states corresponding to a superposition of black hole and white hole states~\cite{Rovelli:2018okm}, with a lifetime that can exceed the age of the Universe due to the combination of the minimal non-zero area eigenvalue and the classical large internal volume.  

The usual history of the early Universe, in which inflation, reheating, radiation and matter-dominated eras follow one another, can be altered in the presence of light PBHs evolving to Planckian remnants.  For instance, PBHs can transiently dominate the energy density of the Universe before their evaporation.  This changes the time when inflationary fluctuations re-enter the Hubble horizon and thereby also changes the PBH mass associated with the fluctuation modes, which in turn impacts the duration of this early matter-dominated era (EMD).  These scenarios also provide an alternative way to produce particles ( including possible DM particles~\cite{Bell:1998jk,Fujita:2014hha,Allahverdi:2017sks,Baldes:2020nuv,Masina:2020xhk,Cheek:2021odj,Cheek:2021cfe}) through the PBH evaporation.  This also impacts the required initial densities in each species in order to lead to a late-time cosmology in agreement with the cosmic microwave background and large-scale structure observations.

Constraints on scenarios with PBH evaporation and DM made of Planckian relics have been considered in several works, for instance in~\cite{Barrau:2021spy} for constraints from remnant stability in specific models and from Big-Bang nucleosynthesis, but without considering the possibility of an early PBH-dominated era, from scalar-induced gravitational waves in~\cite{Trivedi:2025agk} where it is even claimed that Gaussian Planckian relics are excluded by LIGO/Virgo/KAGRA observations, and from the Hawking spectrum of evaporating PBHs in~\cite{Vidotto:2016jqx,Barrau:2018rts}. We highlight that Planckian relics can be considered also as product of PBH formed in a bouncing scenarios ~\cite{Carr:2011hv,Quintin:2016qro,Chen:2022usd,Papanikolaou:2023crz,Frion:2025cyd} where most constraints could be avoided, but we do not consider here.\pagebreak

The goal of this paper is to go beyond previous works by exploring diverse early-Universe cosmologies and, in each case, by determining the required initial PBH mass and abundance for the remnants to constitute the DM or a  fraction of it.  In particular we identify the maximum abundance of remnants today as a function of the initial PBH mass.  In addition, we calculate some observable signatures of these scenarios, e.g. the changes in the number of relativistic degrees of freedom and the presence of a gravitational-wave background induced by the large scalar perturbations at the origin of PBHs and by the Poisson fluctuations in the distribution of these PBHs before their evaporation and the formation of Planckian relics.  This way, we provide new constraints on the parameter space of these scenarios and identify the regions that could be probed with future observations.  {Upon completion of this work, we noticed that similar constraints on the PBH abundance from GWs induced by Poisson fluctuations have been found in~\cite{Domenech:2023mqk} based on simple analytical arguments.  Besides confirming their results, our constraints also include scalar-induced GWs from primordial fluctuations and effects of evaporation on the relativistic degrees of freedom.  We also provide a more detailed and exhaustive analysis of all possible cosmologies with Planckian remnants, combining numerical and analytical calculations, leading to new constraints on remnant stability and including a deep discussion of the phenomenology associated with a particular PBH mass around $10^3$ kg, and of the absence of fine-tuning for the remnant abundance.}

The paper is organized as follows:  In Section~\ref{sec:PBHs and Planckian remnant formation in LQG} we briefly review the production of Planckian remnants from PBHs in LQG.  The procedure used to determine the initial conditions is presented in Section~\ref{sec:Initial abundances} and the resulting cosmologies with different phases are discussed in Section~\ref{sec:PBH cosmologies in LQG}.  Then, constraints on the initial PBH abundance are calculated in Section~\ref{sec:Constraints on the initial PBH fraction} in two regimes, with and without an early PBH-dominated epoch and respectively leading to DM saturation by remnants or radiation saturation by Hawking evaporation products.  The resulting constraints on the remnant abundance are derived in Section~\ref{sec:Constraints on the remnant abundance}.  Other constraints from GW production are obtained in Section~\ref{sec:Observable signatures}.  In Section~\ref{sec:temperature_reheating}, we discuss the relation between the bath temperature and PBH accretion as well as the independence of the reheating temperature reached after PBH evaporation.  We present our conclusions and some perspectives in Section~\ref{sec:Conclusion and perspectives}.  

\section{PBH and Planckian remnant formation in LQG}\label{sec:PBHs and Planckian remnant formation in LQG}

The standard semiclassical approach of black holes treats quantum fields on a fixed classical curved background~\cite{Hawking:1974rv}. This approximation leads to the thermal Hawking radiation of black holes, with an associated temperature $T_\text{\tiny H}\sim 1/m$ where $m$ is the black hole mass. As the black hole loses mass, its temperature rises, leading to a runaway evaporation process. However, the semiclassical approximation fails when the curvature approaches the Planck scale\footnote{Throughout the paper we assume Planck units ($c = G = \hbar = k_B = 1$), unless otherwise specified.} $K\sim 1$, where $K$ is the Kretschmann scalar. LQG provides a non-perturbative, background-independent framework to address this regime. In LQG, the spectra of geometric operators, such as the area operator $A$, are quantized:
\begin{align*}
    A_j \sim \sqrt{j(j+1)}~,
\end{align*}
with half-integer $j$. This leads to the existence of a natural UV cut-off, together with a maximal energy density $\rho_{\rm max}$, which modifies the first Friedmann-Lema\^itre equation and in turn motivates an effective metric describing the internal and external geometry of a BH sufficiently away from the region of Planckian curvature~\cite{Han:2023wxg},
\begin{equation}\label{eq:Schwarzschild LQC}
    \begin{aligned}
        \d s^2 = - \widetilde{f}(r) \d t^2
        + \frac{1}{\widetilde{f}(r)} \d r^2 + r^2 \d\Omega^2~,\\
        \text{with } \widetilde{f}(r) = 1 - \frac{2m}{r} - \frac{3m}{2\pi\rho_\maxi r^4}~.
    \end{aligned}
\end{equation}
This effective geometry removes the BH singularity and replaces it with a bounce into an antitrapped spacetime region. For Eq.~\ref{eq:Schwarzschild LQC} to have a unique asymptotic region, it is necessary to provide also a non-perturbative description of transition in the horizon region. A central prediction of this framework is the possibility of quantum tunnelling from a black hole geometry to a white hole geometry, with a transition probability $P$ exponentially suppressed by the squared mass~\cite{Christodoulou:2018ryl},
\begin{align}
    P \sim e^{-m^2}~.
\end{align}
For a PBH that has evaporated down to the Planck regime ($m\sim 1$), the tunnelling probability becomes of order unity. This mechanism provides a pathway for the formation of remnants. Indeed, the object undergoing the transition is stabilized by its huge interior~\cite{Christodoulou:2014yia} and is in a state that is a superposition of the two geometries (black hole and white hole). These remnants would interact with ordinary matter only through gravitational interactions, making them promising candidates for DM.

The standard semiclassical lifetime of a black hole scales as $\tau_\text{\tiny SC} \sim m_0^3$, where $m_0$ is the initial mass of the BH.\footnote{Possible modifications to this approximation due to the presence of a thermal bath are discussed in Subsec.~\ref{subsec:PBH Reheating and Thermalization (Regime II)}.} In the remnant scenario, the total lifetime $\tau_\tot^{(k)}$ is extended and is bounded by the minimal time necessary for the information in the remnant interior to come out. We adopt a phenomenological parametrization for the latter as:
\begin{align}\label{eq:tau tot}
    \tau_\tot^{(k)} \sim \left( m_0^3 + m_0^{3+k}\right) \sim m_0^{3+k}~.
\end{align}
Theoretical estimates suggest $k\geq1$~\cite{Rovelli:2024sjl,Martin-Dussaud:2025qtr,Bianchi:2026ytw}. If $k$ is  sufficiently large, the remnant acts effectively as a stable particle during the whole history of the Universe.  Constraining $k$ is one primary objective of our analysis. As we will show in Sec.~\ref{sec:Constraints on the remnant abundance}, explaining the totality of DM with remnants originating from very light PBHs (e.g., $m_0 \leq 10^{-3}$ kg) requires high values of $k$ (e.g., $k \ge 14$) in agreement with \cite{Bianchi:2026ytw}.

\section{Initial abundances}\label{sec:Initial abundances}
In this work we assume the following scenario. PBHs form shortly after the end of inflation when a density fluctuation $\delta$ exceeds some threshold $\delta_{\rm c}$ that depends on the equation of state, and re-enters the Hubble horizon. The mass of the resulting PBH is of the order of the horizon mass~\cite{Carr:2009jm},
\begin{align}\label{eq:PBH formation}
    m_\pbh(\Ni) \sim m_H(\Ni) \sim t(\Ni)~,
\end{align}
where $\Ni$ denotes the $e$-fold time (normalized to $N=0$ today) at which PBH formation occurs and $t$ denotes the cosmic time. For simplicity we assume a monochromatic PBH mass distribution. While realistic inflationary models may produce extended mass functions, a monochromatic approximation suffices to identify the distinct phenomenological regimes and leads to clearer constraints on LQG parameters.  We also note that the threshold value may change in loop quantum cosmologies~\cite{Papanikolaou:2023crz}.  The PBH population is therefore fully characterized by two parameters: its initial mass $m_\pbh(\Ni)\equiv \mi$ and its initial abundance $\beta = \rho_\pbh (\Ni) / \rho_{\rm tot}(\Ni)$.  This $\beta$ parameter is a quantity traditionally constrained by various observational probes, often assuming complete evaporation~\cite{Carr:2020gox,LISACosmologyWorkingGroup:2023njw}. A second main goal of our analysis is to re-evaluate these constraints in the presence of stable Planckian remnants.

Even though the total energy density of the Universe at the formation epoch $\rho_\tot(\Ni)$ may be radiation-dominated, we must explicitly track the evolution of all components in order to ensure that the present-day densities match observations. The initial energy density is composed of photons ($\gamma$), neutrinos ($\nu$), baryons ($b$), the cosmological constant ($\Lambda$) and a possible pre-existing cold DM component ($c$) distinct from both PBHs and their subsequent Planckian relics (e.g. weakly interacting massive particles or axions). This a llows us to include in our analysis  scenarios in which PBH remnants constitute only a fraction of the total DM. The initial densities $\rho_A(\Ni)$ of each species $A$ are determined by backward-propagating the present-day observed densities $\rho_A^\obs$ to the formation time $\Ni$, taking into account the standard dilution law $\rho_A \sim e^{-3(1+w_A)\Ni}$, where $w_A$ is the equation of state associated with the species $A$. 
This backward propagation also needs to take into account possible injections due to Hawking radiation. A detailed description is provided in App.~\ref{app:Computation of the alpha parameters} and a \texttt{Mathematica} notebook performing these computations has been publicly released with our paper~\cite{pbh-rem:2025}.

It is worth noting that the relationship between photon and neutrino densities is driven by the standard thermal history of the Universe. Before neutrino decoupling ($T \gtrsim  1$ MeV), neutrinos and photons share a common temperature. After electron-positron annihilation ($T \lesssim 0.5$ MeV), one recovers the standard ratio $\rho_\nu / \rho_\gamma \approx \frac{7}{8} N_\text{eff} (4/11)^{4/3}$. Our modelling explicitly distinguishes cases where PBH evaporation occurs before and after this epoch, because this indicates if the Hawking radiation products thermalize or instead contribute to a \textit{dark radiation} component characterized by $\Delta N_\text{eff}$.  The thermal histories associated with these cases and the resulting constraints will be discussed further in Sec.~\ref{sec:temperature_reheating}.  
\section{PBH cosmologies in LQG}\label{sec:PBH cosmologies in LQG}

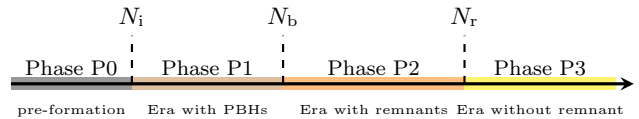
\begin{figure}[h]
    \centering
    \begin{tikzpicture}[>=stealth,thick,scale=0.8]

     \def\Nstart{0}  
     \def\Nii{2.0}      
     \def\Nbh{4.5}       
     \def\Nrov{7.5}    
     \def\Nend{10}    
     
     \fill[black!40] (\Nstart,-0.1) rectangle (\Nii,0.1);
     \node[above] at ($(\Nstart,0)!.5!(\Nii,0)$) {\footnotesize Phase P0};
     \node[below, yshift=-0.15cm, font=\tiny] 
          at ($(\Nstart,0)!.5!(\Nii,0)$) {pre-formation};
     
     \fill[brown!50] (\Nii,-0.1) rectangle (\Nbh,0.1);
     \node[above] at ($(\Nii,0)!.5!(\Nbh,0)$) {\footnotesize Phase P1};
     \node[below, yshift=-0.15cm, font=\footnotesize] 
          at ($(\Nii,0)!.5!(\Nbh,0)$) {\tiny Era with PBHs};

     \fill[orange!50] (\Nbh,-0.1) rectangle (\Nrov,0.1);
     \node[above] at ($(\Nbh,0)!.5!(\Nrov,0)$) {\footnotesize Phase P2};
     \node[below, yshift=-0.15cm, font=\footnotesize] 
          at ($(\Nbh,0)!.5!(\Nrov,0)$) {\tiny Era with remnants};

     \fill[yellow!70] (\Nrov,-0.1) rectangle (\Nend,0.1);
     \node[above] at ($(\Nrov,0)!.5!(\Nend,0)$) {\footnotesize Phase P3};
     \node[below, yshift=-0.15cm, font=\footnotesize] 
          at ($(\Nrov,0)!.5!(\Nend,0)$) {\tiny Era without remnant};

     \draw[->, line width=1.2pt] (\Nstart,0) -- (\Nend+0.3,0);
     
     \draw[dashed] (\Nii,0) -- ++(0,0.8) node[above] {$\Ni$};
     \draw[dashed] (\Nbh,0) -- ++(0,0.8) node[above] {$\Nb$};
     \draw[dashed] (\Nrov,0) -- ++(0,0.8) node[above] {$\Nr$};
     
     \end{tikzpicture}
     
    \caption{The history of the Universe is divided into four distinct phases. Phase 0 (P0): pre-formation era ($N$ $<$ $ \Ni$). Phase 1 (P1): PBH dilution era ($\Ni$ $<$ $N$ $<\Nb$) starting at PBH formation. 
    Phase 2 (P2): remnant era (
    $\Nb$ $<$ $N$ $<$ $\Nr$), triggered by Hawking evaporation of PBHs at $\Nb$. Phase 3 (P3): final era ($N>\Nr$), where remnants have eventually decayed into soft radiation.}
    \label{fig:temp_diagram_phases}
\end{figure}

The evolution of the Universe in the presence of evaporating PBHs leading to Planckian remnants can be complex because of the interplay between multiple fluid components with different equations of state that can transiently dominate the density of the Universe.  To handle this complexity, we have structured the cosmological history into a four-phase model, illustrated in Fig.~\ref{fig:temp_diagram_phases}. This structure allows us to numerically integrate the evolution equations and solve the boundary conditions in the different phases. 

\subsection{Phase 0 (P0): pre-PBH formation}\label{subsec:P0}

As an initial condition, we consider a phase prior to PBH formation, during which the energy density of the Universe is composed of all the species that are not products of the PBH evaporation, including a possible initial cold DM component.  This initial density and relative abundance of each species $A \in \{\gamma, \nu, b, c, \Lambda\}$ must be fixed so that the whole scenario reproduces the abundances observed today, except for neutrinos for which there is only an upper bound.  In order to determine these initial amounts while respecting these conditions, we have used an iterative method. This phase P0 lasts until the  formation of PBHs at $e$-fold time $\Ni$, according to Eq.~\ref{eq:PBH formation}.

\subsection{Phase 1 (P1): era of PBH dilution}\label{subsec:P1}
This phase spans the period from PBH formation ($\Ni$) to the moment of evaporation ($\Nb$), which is effectively instantaneous due to the explosive nature of the Hawking radiation. In this regime, PBHs behave as a pressureless, non-relativistic matter component.  A priori the phase P1 could either be radiation-dominated, matter-dominated ($c$+$b$), PBH dominated or a successive combination of them.   In particular, starting with a radiation-dominated phase, it is possible that the PBH relative fraction grows (because their density dilutes like $a^{-3}$ whereas the radiation density dilutes like $a^{-4}$), until they become the dominant component. Such a scenario would therefore induce a transition to an EMD era during P1.

\subsection{Phase 2 (P2): Era with remnants}

The P2 phase starts at the time of PBH evaporation and remnant formation, and terminates when the remnants decay.   
P2 is triggered by the critical transition between semiclassical physics and non-perturbative (loop) quantum gravity.  The end of the semiclassical phase is marked by the PBH evaporation time $\Nb$. 

We model the evaporation as an instantaneous injection of energy (described thereafter), which is justified given that ($\d m/\d t \propto -1/m^2$).  At the $e$-fold time $\Nb$, the mass density in PBHs is partitioned into two distinct components:
\begin{enumerate}
    \item Planckian remnants: a fraction $\epsilon$ of the PBH mass stabilizes in remnants. In the LQG scenario, the remnant mass is fixed by the Planck scale $m_\text{\tiny P}$ and $\epsilon$ is given by:
    \begin{align}
        \epsilon = \frac{m_\REM}{m_\pbh(\Nb)} = \frac{\left(\sqrt{3 \sqrt3 \gamma_\LQG }/2\right) m_\text{\tiny P}}{m_\pbh(\Ni)}~,
    \end{align}
    where $\gamma_\LQG$ is called the Barbero-Immirzi parameter, assumed here to be of order one without loss of generality.  Compared to the memory burden effect where $\epsilon \sim \mathcal O(1)$ and constant, in LQG with this choice of $\gamma_\LQG$  one has $\epsilon \sim 1/\mi \ll 1$.
    \item Hawking radiation products: the remaining mass fraction $(1 - \epsilon)$ is converted into Standard Model particles and gravitons, but not into beyond Standard Model particles. In order to accurately model this injection, we use the \texttt{BlackHawk} code~\cite{Arbey:2019mbc,Arbey:2021mbl} to compute the spectra for each species that we then integrate over time and frequency. More precisely, \texttt{BlackHawk} produces the secondary spectra of stable particles ($\gamma$, $\nu$, protons $p$, electrons and positrons $e^\pm$ and gravitons $h$) resulting from the decay of primary Hawking products (quarks, gluons, gauge bosons). 
\end{enumerate}

The initial densities in P2 are therefore given by
\begin{align}
    \rho_\REM (\Nb^+)=\epsilon\rho_\pbh(\Nb^-)~,\\\label{eq:Hawking_injection}
    \rho_A(\Nb^+)=(1-\epsilon)\epsilon_A\rho_\pbh(\Nb^-)~.
\end{align}
where $\rho_\REM$ denotes the energy density of the remnants, $\epsilon_A$ are the species-dependent branching ratios derived from \texttt{BlackHawk} and the superscripts ${(-,+)}$ are associated with the $e$-fold times right before or after the transition between P1 and P2.  The remnants then evolve as a stable matter component, potentially being DM.  The photons, protons, electrons and positrons thermalize in the radiation bath, while gravitons free-stream independently.  For neutrinos, they either contribute to the thermal bath or free-stream depending on whether they are injected before or after the neutrino decoupling.
    
\subsection{Phase 3 (P3): Remnant decay and final state}   

Although the primary focus is on stable remnants, our model includes a possible third phase to accommodate scenarios where remnants have a finite lifetime. In this case, they eventually decay into low-energy photons at $N=N_{\rm r}$ that defines the beginning of P3.  By requiring that remnants constitute a stable  fraction of DM, we impose that $N_{\rm r} >0$, which allows us to exclude regions of the parameter space where remnants would have decayed by the present day.  This way we constrain the value of the lifetime parameter $k$, as discussed in Sec.~\ref{sec:Constraints on the remnant abundance}. The lifetime depends on the initial PBH mass as in Eq.~\ref{eq:tau tot},  In reality, observations may allow a small fraction of remnants to decay and this effect could be more complex in the case of extended PBH mass functions.  This will be explored in detail in a future work.

\vspace{2mm}

The evolution through these phases has been simulated using a custom \texttt{Mathematica} code hosted on \texttt{GitHub}~\cite{pbh-rem:2025} and publicly available.  Given the current observed densities $\rho_\text{\tiny DM}^\obs$ $\rho_b^\obs$, $\rho_\Lambda^\obs$ and $\rho_\gamma^\obs$, it iteratively determines the required initial abundance $\beta$ at the formation epoch $\Ni$ and equivalently the initial PBH mass.  The time evolution of all cosmological quantities of interest is then obtained. This method ensures that all the cosmological histories generated are consistent with the $\Lambda$CDM boundary conditions at present time.

\section{Constraints on the initial PBH abundance}\label{sec:Constraints on the initial PBH fraction}

\begin{figure*}[t]
    \centering
    \includegraphics[width=1\textwidth]{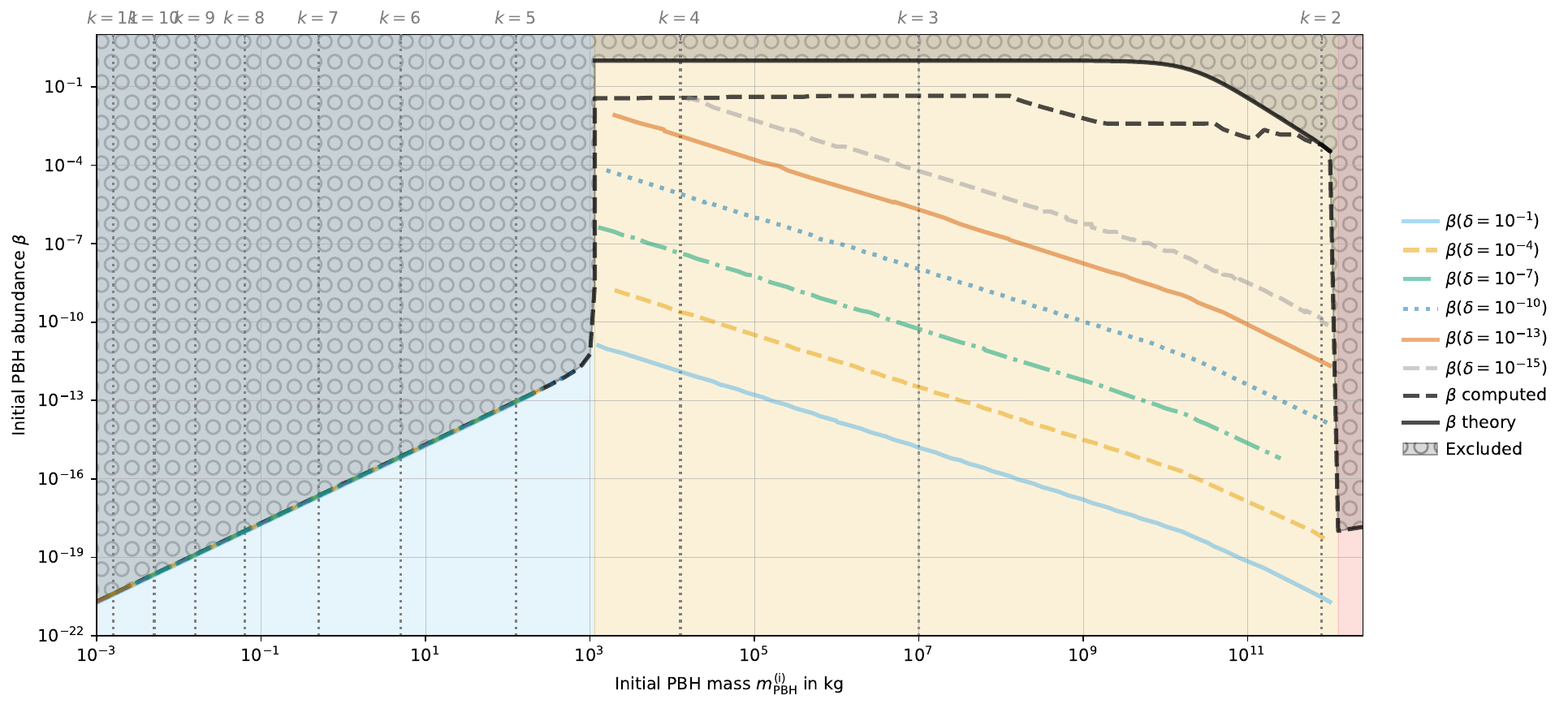}
    \caption{Upper bounds on the initial PBH abundance $\beta$ as a function of the initial mass $\mi$. The parameter space is divided into two main regimes. The blue-shaded background indicates the regime where constraints are driven by the Planckian remnant contribution to DM. Within that mass range, DM is only saturated by remnants ($f_\REM \approx 1$) near the uppermost bounds, whereas for even slightly lower values of $\beta$, the remnant contribution becomes highly subdominant. Similarly, the yellow-shaded background corresponds to the EMD regime where constraints are driven by Hawking evaporation products. The corresponding colored lines represent constant values of $\delta_{\gamma} = 1 - f_{\gamma}$, with the radiation bath being almost entirely generated by PBH evaporation ($f_{\gamma} \approx 1$) only along the upper bounding curves. The continuous black line represents the theoretical maximal value of $\beta$, while the dashed line indicates the numerical limit obtained by our code. The gray region above the theoretical maximal value of $\beta$ is excluded by the overproduction of remnants (Regime I) or radiation (Regime II).  The upper horizontal axis indicates the minimum lifetime parameter $k$ required to ensure the total PBH lifetime $\tau_\tot = \tau_\pbh + \tau_\REM$ exceeds the age of the Universe. }
    \label{fig:results_LQG}
\end{figure*}

Our analysis leads to constraints on the initial PBH abundance $\beta$ as a function of the initial PBH mass $\mi$, shown in Fig.~\ref{fig:results_LQG}. These constraints are driven by the requirement not to overproduce specific cosmic components, as detailed hereafter.

The analytical structure of the model allows for a theoretical derivation of the dependence of the maximal value of $\beta$ with the PBH mass. In the regime where the present-day abundance of a species $A$ is fully saturated by PBHs, their Hawking radiation, or the remnants, the maximum allowed fraction $\beta_A$ is determined by requiring that the parameter $\alpha_A$, controlling the initial pre-existing density of that species, vanishes:
\begin{align}\label{eq:beta_max_cdt}
    \alpha_A\left( \beta, \mi \right) = 0 \quad \longrightarrow \quad \beta_{A} \left( \mi \right)~.
\end{align}
The explicit computation of these coefficients, along with the analytical derivation of $\beta_{A}$, is detailed in App.~\ref{app:Computation of the alpha parameters}.

We identify two distinct regimes governing these constraints in the LQG scenario: one where the DM component is dominated by Planckian remnants, and one where the radiation is dominated by the Hawking evaporation products.  In the latter case, we identify the maximal contribution of remnants to the DM.

\subsection{Regime I: DM saturated by remnants}\label{subsec:RI}
\begin{figure*}
    \centering
    \text{(a) PBHs with an initial mass of $\mi = 10^2$ kg and $\beta \sim 10^{-13}$.}
    \includegraphics[width=.85\linewidth]{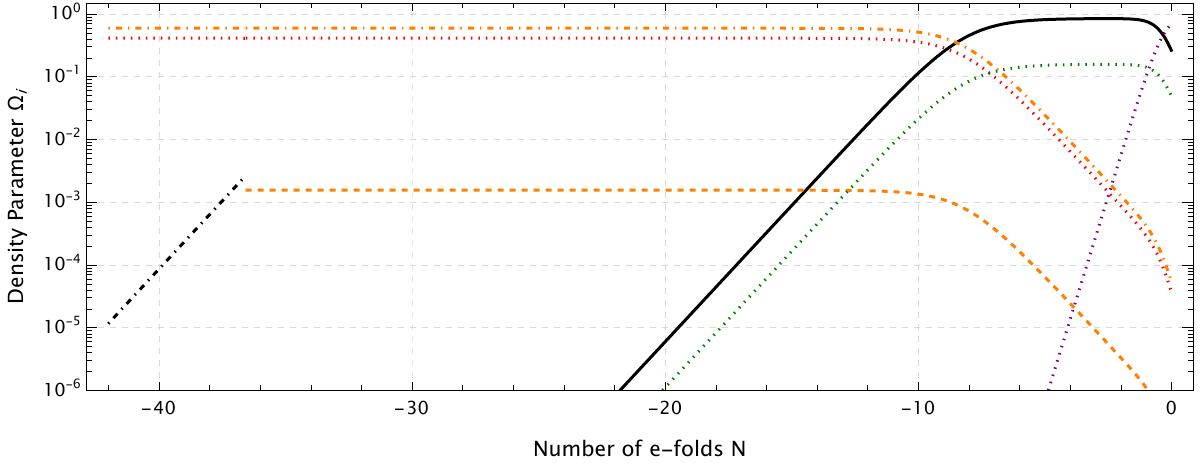}
         
    \centering
    \text{(b) PBHs with an initial mass of $\mi \approx 1.14\times10^3$ kg and $\beta \sim 10^{-4}$.}
    \includegraphics[width=.85\linewidth]{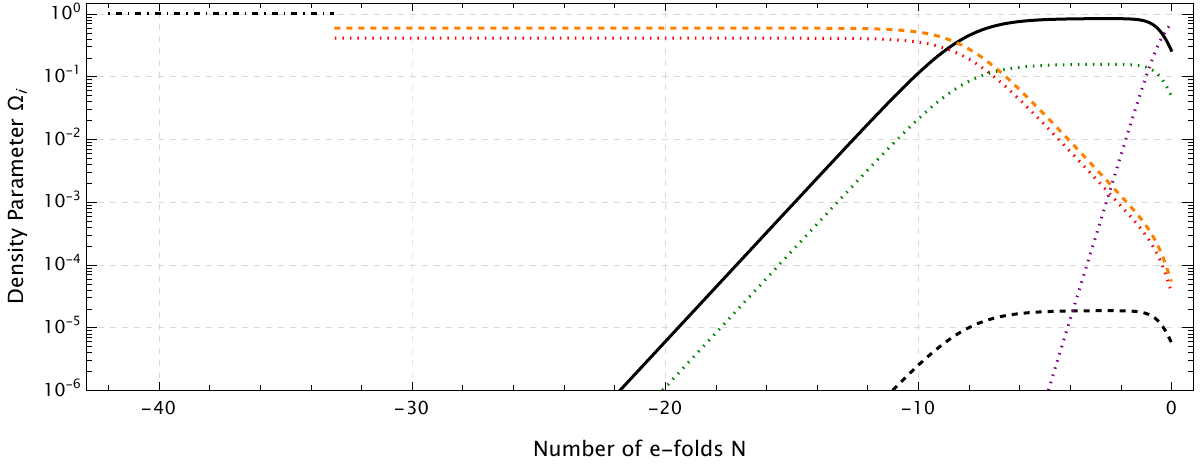}
        
    \centering
    \text{(c) PBHs with an initial mass of $\mi = 10^4$ kg and $\beta \sim 10^{-2}$.}
    \includegraphics[width=.85\linewidth]{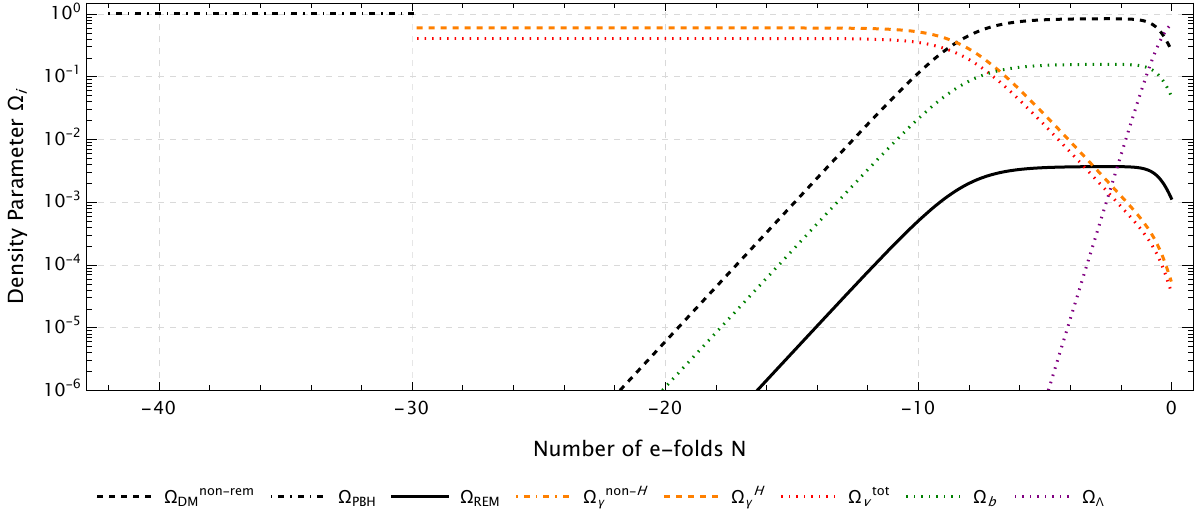}

    \caption{Time evolution of the relative densities of DM excluding remnants $\Omega\indices{_\DM^{\text{non-rem}}}$, PBHs $\Omega_\pbh$, remnants $\Omega_\REM$, initial photons $\Omega\indices{_\gamma^{\text{non-H}}}$, Hawking photons $\Omega\indices{_\gamma^{\text{H}}}$, total neutrinos $\Omega\indices{_\nu^\tot}$, baryons $\Omega_b$ and cosmological constant $\Omega_\Lambda$ as a function of the number of $e$-folds $N$, for different initial PBH masses $\mi$. In the first panel (a), the Universe begins in a standard radiation-dominated era. PBHs evaporate early ($\Nb \approx -35$) before they can dominate the energy density. The resulting Planckian remnants' relative density grows to eventually generate a standard matter-dominated era and saturate the observed DM density today. In the last panel (c), PBHs dominate the energy density early on, creating an EMD era. Upon evaporation ($\Nb \approx -30$), the Universe reheats via Hawking radiation, which saturates the observed photon density. Consequently, the remnant abundance is strictly constrained, limiting its contribution to DM to a fraction $f \sim 10^{-3}$. The middle panel (b) shows the transition between these two regimes. For this specific initial mass $\mi =\mss \approx 1.14\times10^3$ kg, the PBH products explain both the totality of the DM through their Planckian remnants and the totality of the photon density through their Hawking products.
    }
    \label{fig:time_evolution}
\end{figure*}

For very light PBHs, with masses between $10^{-4}$ kg and $10^3$ kg, remnants are viable candidates to account for all the DM.  An example of the time evolution of  relative densities for the different species, for $\mi = 10^2$ kg, is shown in the first panel (a) of Fig.~\ref{fig:time_evolution}.  

In that mass range, the PBH mass is sufficiently low for their lifetime to be short, in such a way that the fraction of mass lost to radiation through Hawking evaporation $(1-\epsilon)$ does not saturate the total radiation component.  
The initial fraction $\beta$ is therefore constrained by the fact that remnants cannot exceed the DM density today.  Their relative contribution to the density of the Universe grows linearly with the scale factor in the radiation era and exactly matches $\Omega_\DM^\obs$ today. 

The resulting constraint on $\beta$ goes like $\beta_{\rm c} \propto(\mi)^{3/2}$ and reaches $10^{-12}$ for PBHs of about $10^3$ kg.  This behaviour can be explained by solving Eqs.~\ref{eq:beta_max_cdt} in App.~\ref{app:Computation of the alpha parameters} when requiring that DM is solely made of remnants, which implies that the pre-existing cold DM density vanishes. One then gets (see Eq.~\ref{eq:beta_c} in App.~\ref{app:Computation of the alpha parameters}):
\begin{equation}
\beta_{\rm c} \sim e^{\Ni}/\epsilon \sim \mi^{1/2} \mi \propto\mi^{3/2}~.
\end{equation}

Compared to the case usually considered with no remnant, this scenario opens a viable window where very light PBHs evolve into stable Planckian remnants that account for the whole DM.  

\subsection{Regime II: photon saturation}\label{subsec:RII}

For larger PBH masses, above $10^3$ kg, a much larger fraction of the initial mass $(1-\epsilon)$ is released into Hawking evaporation products, and at the same time, the longer lifetime dilutes more the pre-existing radiation.  As a result, at the time when the PBHs evaporate, the Hawking products become the dominant component of the Universe.  When requiring to end up with the correct radiation and matter abundances today, one gets that $\beta$ is of order one and that PBHs quickly lead to an EMD era.  In this regime, almost all the radiation content originates from the PBH evaporation process.  If the PBHs had an extended mass function, this constraint and regime will probably differ in a non-trivial way and this is left for a future analysis.   

Above $10^3$ kg, the photon density today originates from the PBH evaporation products but the maximal DM fraction made of Planckian remnants decreases like (following Eq.~\ref{eq:beta_gamma}) $(\mi)^{-9/2}$, down to $10^{-22}$ for $\mi \sim 10^{12}$ kg above which PBHs have a longer lifetime than the age of the Universe.  Again, this is explained by the calculation in App.~\ref{app:Computation of the alpha parameters}, when requiring that the initial photon density vanishes.

Consequently, the maximum fraction of DM that can be constituted by remnants $f_\REM$ drops precipitously as $\mi$ increases. For $\mi \gtrsim 10^5$ kg, remnants can only explain a small percentage of DM.

\section{Constraints from remnant abundance and evaporation products}\label{sec:Constraints on the remnant abundance}

Combining the identified regimes, we establish the \textit{viable mass window} where Planckian relics account for the totality of DM ($f_\REM = 1$) without violating background radiation constraints:
\begin{equation}
    \mi \in [10^{-4},10^{3}]\text{ kg}~.
\end{equation}
The lower bound is set by the constraint on the scalar-to-tensor ratio, which limits the scale of inflation $H_{\rm inf}$. Allowing remnants to be a sub-dominant but still sizeable DM component ($f_\REM \gtrsim 10^{-5}$) extends the window to $\mi \in [10^{-4},10^{5}]$ kg. For larger masses, the photon saturation resulting from PBH evaporation leads to remnants being cosmologically negligible.

Interestingly, a cosmological ``sweet spot'' occurs at the transition between the two identified regimes ($\mi \approx 1.14 \times 10^3 \text{ kg} \equiv \mss$), where PBHs both generate the observed DM and reheat the Universe through their evaporation. This scenario is achievable for a wide range of initial PBH abundances from $\beta \sim 10^{-11}$ up to $\beta \sim \mathcal{O}(1)$.

\subsection{Analytical estimate of the sweet spot mass}

The sweet spot mass can be estimated solely by requiring that the DM is entirely constituted by remnants and that the cosmic radiation bath is exclusively generated by Hawking evaporation. Expressing $\mi$ as a function of the observed ratio $R_\obs \equiv \rho_{\DM}^\obs/\rho_\text{rad}^\obs$ and the radiation density $\rho_\text{rad}^\obs$, we assume a sufficiently large $\beta$ to lead to an EMD era where $a \sim t^{2/3}$ and $H_\text{b} \approx 2/(3\tau_\pbh)$. The Friedmann-Lema\^itre equation gives:
\begin{equation}
    \rgh(\Nb) \approx \frac{3}{8\pi} H_\text{b}^2 \approx \frac{1}{6\pi\cdot 10^4} \frac{1}{(\mi)^6}~,
\end{equation}
using $\tau_\pbh \approx 10^2\, (\mi)^3$. Matching the observed ratio $R_\obs = \rho_\REM / \rgh \approx \epsilon e^{-\Nb}$ yields $e^{\Nb} = \epsilon/R_\obs$. Consequently:
\begin{align}
    \rho_\text{rad}^\obs 
        &\approx \rgh (\Nb) e^{4\Nb} \approx \frac{1}{6\pi\cdot 10^4}\frac{\epsilon^4}{{R_\obs}^4}\frac{1}{(\mi)^6}~.
\end{align}
With $\epsilon = 1/\mi$, we find $\mss \approx 5~\times~10^{10}\,\approx 1.14 \times 10^3$ kg. Exploration of alternative scenarios proposing other types of remnants, such as the Memory Burden framework ($\epsilon \approx 1/2$), will be studied in an upcoming work.

\subsection{Radiation from Hawking evaporation}

\label{subsec:Genericity of beta in regime II}

In the photon-saturated regime (regime II), the initial fraction $\beta$ is highly degenerate with the duration of the EMD era, offering a remarkably non-fine-tuned cosmological scenario. Because the pre-existing background is exponentially diluted during the PBH-dominated phase, any macroscopic initial PBH abundance inevitably leads to an attractor-like state where Hawking evaporation accounts for the entirety of the present-day radiation budget ($f_\gamma \to 1$).

Importantly, this mechanism does not strictly require an initial radiation bath. It applies equally well to any pre-existing background that redshifts faster than pressureless matter, such as an inflaton field oscillating in a quartic potential $V(\phi) \propto \phi^4$. Assuming for simplicity a background redshifting like radiation at formation $\Ni$, the initial ratio of PBHs to the pre-existing density $\rho_{\rm bg}^{\rm (i)}$ is $\rho_\pbh(\Ni)/\rho_{\rm bg}^{\rm (i)}(\Ni) \approx \beta/(1-\beta)$. Because PBHs behave as pressureless matter, this relative energy density grows by a factor $e^{\Nb - \Ni}$ until the evaporation epoch $\Nb$. At $\Nb$, the Hawking evaporation injects a radiation density $\rho_\gamma^{\rm H}(\Nb) \approx \rho_\pbh(\Nb)$. The ratio of injected radiation to the pre-existing background is governed by the deficit parameter $\delta_\gamma = 1 - f_\gamma$:
\begin{equation}\label{eq:fraction_evap_2}
    \frac{\rho_\gamma^{\rm H}(\Nb)}{\rho_{\rm bg}^{\rm (i)}(\Nb)}
    = \frac{1-\delta_\gamma}{\delta_\gamma}
    \approx \frac{\beta}{1-\beta} e^{\Nb - \Ni} ~.
\end{equation}
Inverting this relation provides an analytical approximation for $\beta$ (see Eq.~\ref{eq:beta_gamma_delta} for the full analytical formula):
\begin{equation}\label{eq:beta_analytical}
    \beta \approx \left( 1 + \frac{\delta_\gamma}{1-\delta_\gamma} e^{\Nb -\Ni} \right)^{-1} \approx \delta_\gamma^{-1} {\rm e}^{-(\Nb - \Ni) }~.
\end{equation}

This explicitly highlights a degeneracy: a larger initial fraction $\beta$ triggers the EMD era earlier, leading to a stronger exponential suppression factor $e^{-(\Nb - \Ni)}$ for the initial background.  Even though $\delta_\gamma$ is very small, this suppression factor pushes $\beta $ to values that can be much smaller than one.  For instance, in Fig.~\ref{fig:results_LQG} one can see that the obtained value of $\delta_\gamma$ is much smaller than one in Regime II over a wide range of $\beta$ values.  In particular, at the sweet spot mass, imposing $\delta_\gamma < 0.01$ (i.e. less than 1\% of radiation not coming from Hawking products), leads to the viable range $10^{-10} \lesssim \beta \lesssim 1$.

We will further come back later to the genericity of the scenario with respect to the initial PBH abundance, in the case of PBH formation from an initial radiation component, and we will show that $\delta_\gamma$ is in fact independent of $\beta$ in the regime of interest.  In the particular case of the sweet spot, we will additionally show why $f_\REM = 1$ is obtained independently of $\beta$.  

Note that simulating this pure attractor state ($\delta_\gamma \to 0$) introduces numerical challenges, as the machine precision of the solver imposes an artificial lower bound on this parameter, an artifact that we more rigorously address in App.~\ref{app:Numerical and temperature}. Therefore, physical constraints dictate that any value of $\beta$ capable of inducing an EMD era represents a viable, non-fine-tuned solution, rendering the final cosmological observables insensitive to the exact initial conditions at $\Ni$. 

\subsection{Stability constraints}

The viability of these scenarios depends on the remnant stability, which is governed by the lifetime parameter $k$. In this work, we derive a new phenomenological lower bound on $k$ by requiring that remnants survive until the present day to constitute the totality (or a part of) the DM. 

For the cosmological sweet spot mass $\mss$, the remnant lifetime requires $k \ge 4$. This constraint becomes more stringent for lighter PBHs, reaching up to $k \ge 14$ at the extreme lower end of our mass window. Higher value of $k$ are favoured by current estimations \cite{Bianchi:2026ytw} where the lifetime parameter depends on the mass with which the original black hole formed: $k=k(m_0)$. In this framework, one finds, for example, $k \sim 10^8$ for $m_0 = 10^{-3}$ kg and $k \sim 10^{20}$ for $m_0 = 10^3$ kg. 
Notice that earlier estimates suggested lower, constant values, with $k\geq1$ expected from information release constraints~\cite{Preskill:1992tc,Rovelli:2024sjl} and $k=2$ derived from the internal white hole dynamics~\cite{Martin-Dussaud:2025qtr}, prompting in the past some hasty dismissal of cosmological scenarios with remnants from ultralight PBH.

\section{Other observational constraints}\label{sec:Observable signatures}

The presence of light PBHs in the early Universe, even if they have evaporated to form remnants by today, should have left detectable imprints, in particular in the form of a GW background radiation.  GWs are produced by three mechanisms:  i) scalar-induced GWs due to the existence of large scalar fluctuations leading to PBH formation, ii) in the case of an EMD epoch, Poisson fluctuations in the PBH distribution grow and source GWs, iii) when PBHs evaporate they emit gravitons that contribute to the cosmological GW background.  The latter mechanism is shortly mentioned in the next section, however we find that for the cases of interest the GW background is far from being detectable with the current techniques. In the case of scalar-induced GWs, one should also note that they are amplified by the existence of a PBH-dominated  era followed by a sudden  transition from EMD to RD era, an effect referred to as the Poltergeist mechanism and studied in~\cite{Zeng:2025ecx,Inomata:2025wiv}.  

\subsection{Regime I:  DM saturated by remnants}

In the regime where Planckian remnants saturate DM and for which there is no PBH-dominated epoch, GWs only arise from scalar-induced tensor fluctuations.  This GW background is sourced at second order in perturbation theory by the large scalar fluctuations at the origin of PBH formation.  There is an extensive literature on the topic, so we provide below the final formula to compute this GW background for a sharp peak in the primordial power spectrum giving rise to an (almost) monochromatic PBH mass distribution.  In this case, the GW spectrum can be calculated analytically as~\cite{Kohri:2018awv}

\begin{eqnarray}
\Omega_\GW(k,t_k) & =& \,\frac{3 \mathcal{P}_{\rm p}^2 }{64  } \frac{k^2}{k_{\rm p}^2} \left[\frac{4-(k/k_{\rm p})^2}{4}\right]^2\left( 3 \frac{k^2}{k_{\rm p}^2} - 2 \right)^2 \nonumber\\
& \times & \left\{ \pi^2 \left( 3 \frac{k^2}{k_{\rm p}^2} - 2 \right)^2 \Theta \left( 2 \sqrt 3 - 3 \frac{k^2}{k_{\rm p}^2} \right) \right. \nonumber\\
&+ & \left. \left[ 4+\left(3 \frac{k^2}{k_{\rm p}^2} - 2   \right) \ln \left| 1 - \frac{4 k_{\rm p}^2}{3 k^2}\right|  \right]^2 \right\} \nonumber \\
& \times & \Theta \left( 2 -  \frac{k}{k_{\rm p}} \right)~, \label{eq:OmGWSI}
\end{eqnarray}

where $k_{\rm p} $ and $\mathcal P_{\rm p}$ are the position and amplitude of the peak in the primordial curvature power spectrum.  For $k\ll k_{\rm p}$, the spectrum is suppressed, there is a peak at $k\sim k_{\rm p}$ and the spectrum ends abruptly at $k = 2 k_{\rm p}$.  One can then relate it to the GW spectrum today~\cite{Kohri:2018awv},

\be
\ogwo(k) \approx \frac{a_{\rm eq}}{a_0}  \Omega_\GW(k,t_k),
\ee

when the modes of interest re-enter in the Hubble horizon deep in the radiation era and by considering that GWs are diluted like radiation after the matter-radiation equality.  Finally, the peak amplitude can be related to $\beta$ using the common approximation based on the Press-Schechter formalism~\cite{Harada:2013epa}, 
\be\label{eq:Press-Schechter}
\beta =  {\rm erfc} \left( \frac{\zeta_{\rm c}}{\sqrt {2 \mathcal P_{\rm p}}} \right) \approx \frac{\sqrt{2 \mathcal P_{\rm p} }}{ \sqrt \pi \zeta_{\rm c}} {\rm e}^{- \frac{\zeta_{\rm c}^2}{2 \mathcal P_{\rm p}}}~,
\ee
where $\zeta_{\rm c} \approx 0.68$ is the collapse threshold of curvature fluctuations. The last approximation is valid as long as $\mathcal P_{\rm p} \ll \zeta_{\rm c}^2 $, which is usually the case for $\beta \ll 1$ expected in Regime I.  More accurate relations can be used for the calculation of $\beta$ in terms of a density threshold, the compaction function associated with the density profile, with a window function and including a scaling relation for the threshold, but overall the impact of such a refined calculation can be compensated by a slightly different value of the critical threshold value $\zeta_{\rm c}$ (as suggested to be the case in bouncing cosmologies~\cite{Papanikolaou:2023crz,Papanikolaou:2024fzf}), so we prefer to rely on the approximate formalism in this exploratory work in order to maintain and better understand the dependencies of the parameters.  

\begin{figure}
    \centering
    \includegraphics[width=\linewidth]{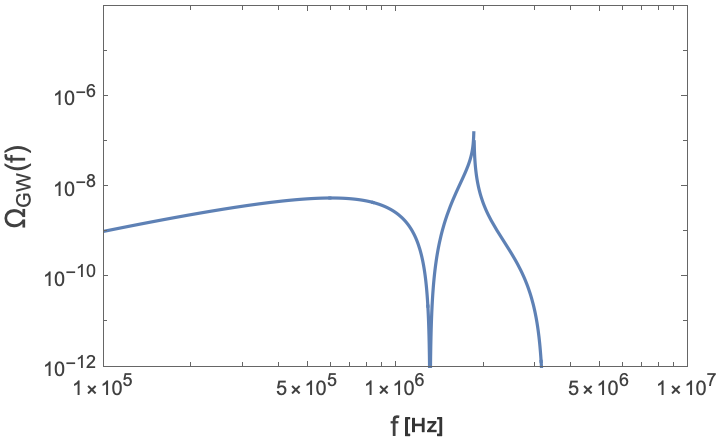}
    \caption{Scalar-induced GW spectrum today for a PBH mass of 10 kg and initial abundance $\beta = 6.47\times 10^{-14} $ leading to DM made of Planckian relics, which corresponds to a peak $\mathcal P_{\rm p} = 7.22 \times 10^{-3}$ at the wavenumber $k_{\rm p} \approx 10^{21} \, {\rm Mpc}^{-1}$ in the primordial curvature power spectrum, leading to a peak at GW frequency $f \approx 2 \times 10^6 {\rm Hz}$.  
    }
    \label{fig:omgw_SI}
\end{figure}

We have represented in Fig.~\ref{fig:omgw_SI} the expected scalar-induced GW spectrum for an initial PBH mass of $10$ kg, which corresponds to $k_{\rm p} \simeq 10^{21} {\rm Mpc}^{-1}$, resulting in a peak at GW frequencies in the MHz range, far above what can be probed with LIGO/Virgo/KAGRA\footnote{We point out that the frequency associated with low-mass PBHs can be well above the range probed by LIGO/Virgo/KAGRA, which seems to have been overlooked in~\cite{Trivedi:2025agk} where it was claimed that LIGO/Virgo/KAGRA limits on scalar-induced GWs exclude Planckian relics for Gaussian primordial fluctuations.}, but in the range of high-frequency GW detectors based on microwave cavities~\cite{Herman:2022fau,Herman:2020wao,Berlin:2021txa}, even if it is challenging to reach the required sensitivity.  This GW background  could also be probed by means that are sensitive to the integrated background spectrum.  For a given value of $\beta$, it is interesting to note that the integrated GW spectrum does not depend on the PBH mass and a numerical integration of Eq.~\ref{eq:OmGWSI} gives
\be
\Omega_\GW^{\rm SI} \approx 3.1 \times 10^{-4} \times \mathcal P_{\rm p}^2~,
\ee
where $\mathcal P_{\rm p}$ is obtained for a given $\beta$ by inverting Eq.~\ref{eq:Press-Schechter}, so it also depends on the critical threshold value.  One gets for instance $\mathcal P_{\rm p} \approx 0.008 $ for $\beta = 6.47 \times 10^{-14}$, which is expected if DM is made of remnants from PBHs with a mass of 10 kg, giving $\Omega_\GW^{\rm SI} \approx 2.0 \times 10^{-8}.$ The existence of such GW background acts like a dark radiation component, which effectively changes the value of $\Neff$.  
The basic relation linking $\Neff$ to an early-seeded GW background can be found in \cite{Cang:2022jyc} and comes from the relation
\be
\rho_\gamma + \rho_\nu + \rho_\GW = \frac{\pi^2}{15}T_\gamma^4 + \frac{7 \pi^2}{120} T_\nu^4 \Neff~,
\ee
where $T_\gamma = T_{\rm CMB} (1+z) $ and $T_\nu = (4/11)^{1/3} T_\gamma$, with $T_{\rm CMB} = 2.728$ K being the CMB temperature today.  One usually decomposes the effective number of relativistic degrees of freedom $\Neff$ into the value expected by the Standard Model and a variation due to exotic components, $\Neff = 3.046 + \Delta \Neff$.  One then relates $\Delta \Neff$ to $\Omega_{\GW}$, integrated over all frequencies and for sources prior to the BBN, as
\be
\Delta \Neff = 8.3 \times 10^4 \Omega_{\GW}~.
\ee
The current constraint is obtained by combining the Planck, BBN and BAO observations, $\Delta \Neff < 0.15$ at 95\% \cite{Planck:2018vyg}, leading to
\be\label{eq:omegaGW_max}
\ogwo \lesssim 1.8 \times 10^{-6}~,
\ee
that can be translated to a constraint $\beta \lesssim 0.14 $ assuming $\zeta_{\rm c} = 0.68$.  
This limit should improve in the future, possibly down to $\Delta \Neff \lesssim 0.03$ for Euclid combined with future CMB observations of Stage IV or LiteBIRD or with 21cm observations of the SKA~\cite{Sakr:2022ans}, constraining $\beta \lesssim 2 \times 10^{-4}$.  This is still far above the expected value in Regime I.  One would actually need to probe $\Delta N_{\rm eff} \lesssim  2 \times 10^{-3}$ to set relevant constraints on Regime 1. 

\subsection{Regime II:  Radiation saturated by Hawking radiation}

\begin{figure*}
    \centering
    \includegraphics[width=\linewidth]{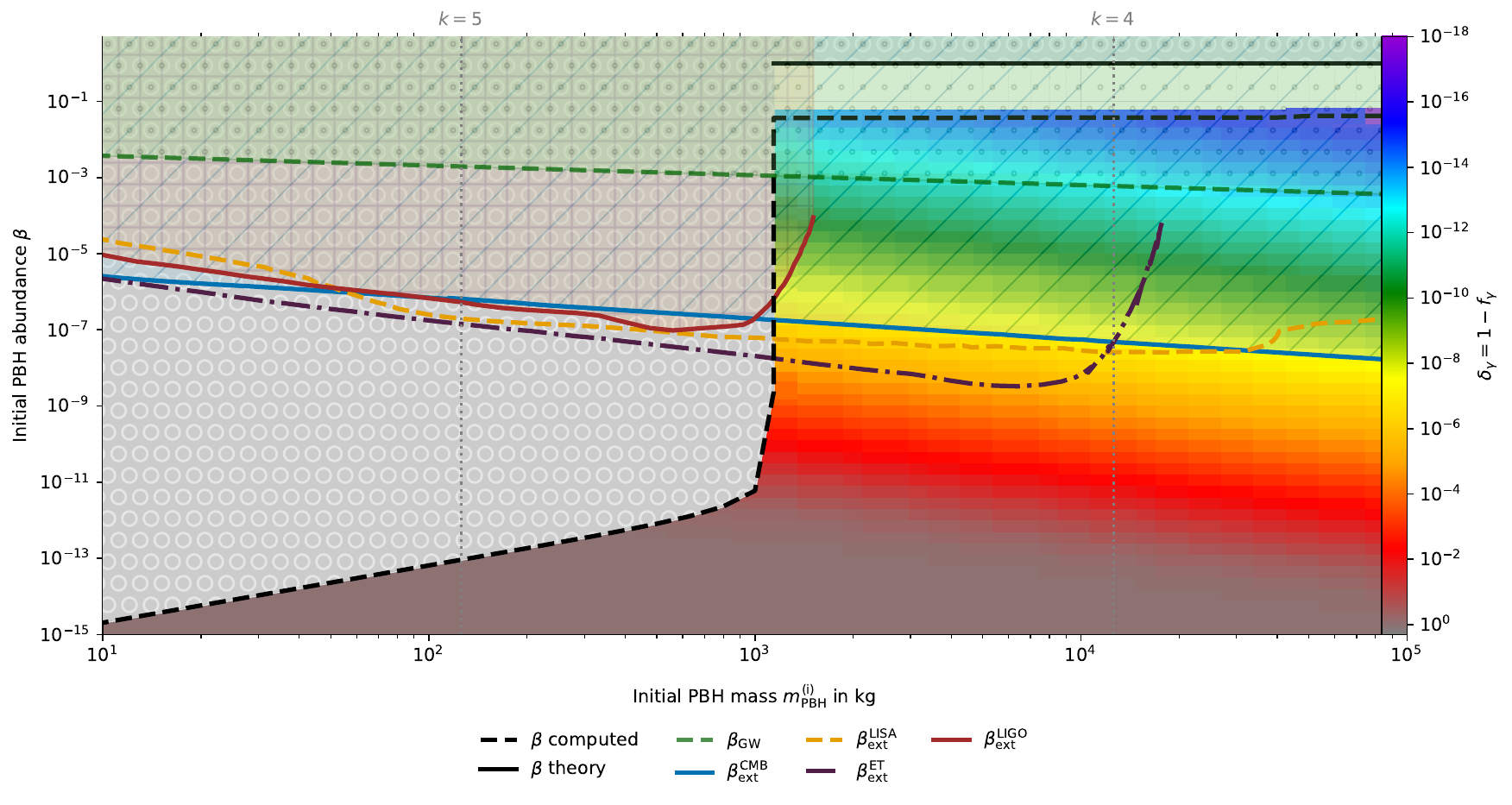}
    \caption{
    Zoom of the parameter space $(\mi,\beta)$ around the ``sweet spot'' mass $\mss \approx 1.14 \times 10^3$ kg between Regime I and Regime II. The heatmap displays the value of the deficit parameter $\delta_\gamma = 1-f_\gamma$ for each allowed value of $(\mi,\beta)$. The continuous black line shows the theoretical maximal value of $\beta$, while the dashed one shows the maximal value obtained numerically. The shaded and hatched regions show where the parameter space is either already excluded by observations (CMB, LIGO/Virgo/KAGRA, and GWs from evaporation) or within the reach of future GW observatories (LISA, ET), for GWs from Poisson fluctuations ($\bgw$) and including the Poltergeist mechanism ($\beta_{\rm ext}$).
    }
    \label{fig:results_sweet}
\end{figure*}

In the regime where PBHs transiently dominate the density of the Universe and their Hawking evaporation products saturate the radiation density, we expect a double-peaked GW background.  A relation between the relative energy density of GW at evaporation time $\ogwe$ and the density of PBHs at formation $\bgw$ was derived in~\cite{Papanikolaou:2020qtd} for a monochromatic distribution. Approximating $\ogwo \approx \ogwe (a_{\rm eq}/ a_0) \approx \ogwe \oro$, it reads:
\begin{align}\label{eq:omega_beta}
    \ogwo = \oro \,\mu \left[ \kappa - \ln(\bgw)\right] \bgw^{16/3}~,
\end{align}
where
\begin{align*}
    \mu &= \frac{1}{16}\left(\frac{45}{2}\mi\right)^{4/3}\left(\frac{g_\text{eff}}{100}\right)^{-2/3}~, \\
    \kappa &= \frac{4}{3\sqrt{5}\pi} +\frac{3}{2}\ln{(2)} ~,
\end{align*} 
with $g_\text{eff}$ given by the effective number of degrees of freedom.
They also derived an upper bound on $\beta$, denoted $\bgw$, by requiring that such GWs do not dominate the density of the Universe, $\ogwe<1$ (or, equivalently, that $\ogwo < \oro$).  Here, we extend their calculation in order to set a limit from the constraints on the number of extra relativistic degrees of freedom $\Delta \Neff$ that also contains the GWs produced before BBN.  Solving Eq.~\ref{eq:omega_beta} is done through the use of the Lambert function $W$ that solves $Y = X e^{X}$:
\begin{subequations}\label{eq:beta_max_GW}
\begin{align}
    \bgw &= \exp{\frac{3}{16} W_{(-)}[Y] + \kappa}\\
        &=\left(-\frac{3}{16}\frac{\oro}{\ogwo}\mu \,W_{(-)}[Y]\right)^{-3/16}~,
\end{align}
where
\begin{align}
    Y = -\frac{16}{3} \frac{\ogwo}{\mu\, \oro} e^{-16 \kappa/3}~.
\end{align}
The argument of the Lambert function $Y$ runs from $-10^{-9}$ to $-10^{-29}$ over the range of PBH masses considered in this study.  Therefore we can consider the $-1$ branch of the Lambert function.
Results from~\cite{Papanikolaou:2020qtd} are recovered  by fixing  $\ogwo=\oro$. One can lower $\bgw$ further by using constraints derived for $\Delta \Neff$~\cite{Caprini:2018mtu} using Eq.~\ref{eq:omegaGW_max} where we took $\Delta \Neff = 0.15$~\cite{Planck:2018vyg} and $h=0.67$. The corresponding mass fraction $\bgw$ is shown in Fig.~\ref{fig:results_sweet}.
\end{subequations}

The second peak comes from the scalar-induced GWs amplified by the Poltergeist mechanism expected for an EMD with a monochromatic PBH distribution. Constraints have been obtained for a variety of probes in ~\cite{Bhaumik:2022zdd}, considering the full double-peak spectrum, but without considering that PBHs leave Planckian remnants.  Nevertheless, the existence of remnants do not impact these constraints because in the regime of transient PBH domination, they are always subdominant in the subsequent cosmic history, as discussed in Sec.~\ref{sec:Constraints on the initial PBH fraction}.  We have reproduced these constraints in Fig.~\ref{fig:results_sweet}, specifically zooming in on the transition region around the ``sweet spot'' mass $\mss$.

\section{Temperature and Reheating}
\label{sec:temperature_reheating}

The thermal history of the Universe is profoundly altered by the presence of light PBHs, either through their interaction with the primordial plasma or via their subsequent evaporation. The phenomenological implications depend intrinsically on the PBH mass and initial abundance.

\subsection{Thermal Evolution and Accretion (Regime I)}

\begin{figure}
    \centering
    \includegraphics[width=\linewidth]{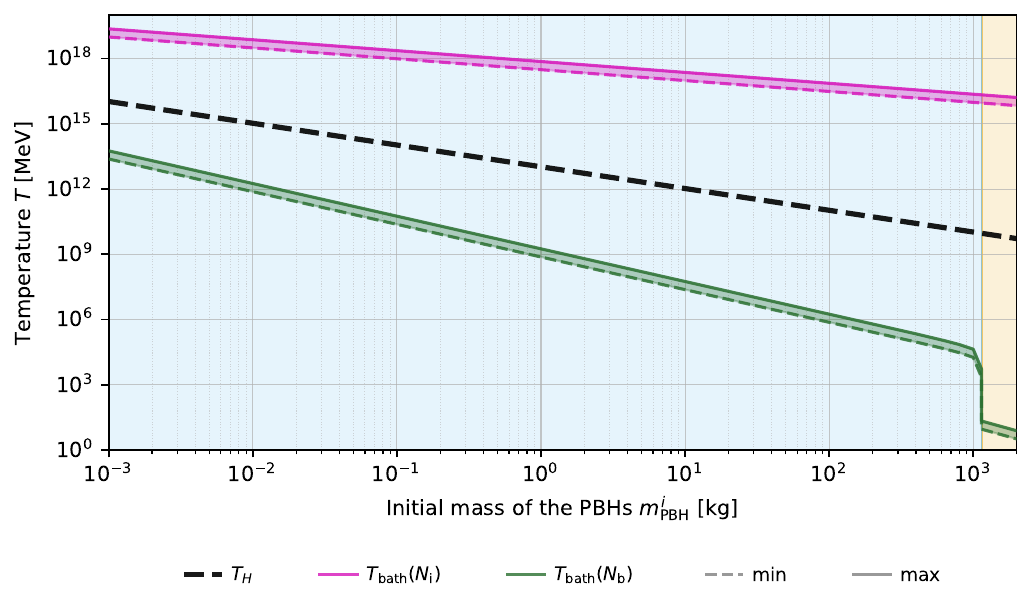}
    \caption{Temperature of the thermal bath at $N=\Ni$ and $N=\Nb$, compared to the Hawking emission temperature.}
    \label{fig:results_reg1_temperature}
\end{figure}

For light PBHs, the early stages of evolution are characterized by a competition between Hawking emission and accretion from the thermal bath. Since PBHs have a negative heat capacity ($C = \partial m / \partial T \sim -m^2 < 0$), they cannot maintain stable thermal equilibrium with their surroundings. At the time of PBH formation, as illustrated in Fig.~\ref{fig:results_reg1_temperature}, the Hawking temperature $T_\text{\tiny H} \sim 1/m$ is strictly lower than the bath temperature $T_{\rm bath}$, which means that accretion should dominate. The standard accretion rate scales as $\dot{m} \sim m^2 T_{\rm bath}^4$.

However, recent semi-classical treatments using the ``Thermofield Double'' formalism suggest that a thermal bath actually enhances the evaporation rate, an effect initially studied in the absence of accretion \cite{Kalita:2025foa}. When simultaneously accounting for both the enhanced evaporation rate and standard accretion, tracking the full dynamical evolution reveals that the net effect is typically negligible \cite{Chaudhuri:2026zyx} for PBHs because the Hubble expansion quickly dilutes the surrounding radiation, making $T_{\rm bath}$ drop drastically. Consequently, even though the initial bath is hotter than the Hawking temperature, it does not significantly change the semi-classical lifetime of PBHs nor our results.

\subsection{Reheating, Thermalization (Regime II) \\and independence from $\beta$}\label{subsec:PBH Reheating and Thermalization (Regime II)}

In scenarios where PBHs transiently dominate the energy density (Regime II), their evaporation acts as a novel reheating mechanism \cite{Garcia-Bellido:1996mdl,RiajulHaque:2023cqe,Klipfel:2026bar}. Crucially, as we show below, the resulting reheating temperature -- and thus the final radiation energy density -- is independent of the initial PBH abundance $\beta$, provided $\beta$ is sufficiently large to induce an EMD era\footnote{During the final stages of this work, we noted that the authors of~\cite{RiajulHaque:2023cqe} already derived this independence of the reheating temperature from the initial PBH abundance, as well as similar EMD cosmological scenarios driven by PBHs, albeit without considering stable Planckian remnants.}.

To understand the independence of $\rgh$ from $\beta$, the early Universe history can be split into a radiation-dominated (RD) era from $\Ni$ to $\Nd$ (when $\rho_\pbh = \rho^{(\rm i)}_{\rm bg}$), and an EMD era from $\Nd$ to $\Nb$. At $\Nd$, we have:
\begin{equation} \label{eq:eq_rad_mat}
    \beta \rho_\tot (\Ni) e^{-3(\Nd - \Ni)} \approx \rho_\tot (\Ni) e^{-4(\Nd - \Ni)}~,
\end{equation}
meaning that during the RD era $e^{\Nd - \Ni} = \beta^{-1}$. In this era, the Hubble rate evolves such that $H_\dom^2 = H_\text{i}^2 \beta^4$, while in the EMD era, it evolves such that $H_{\rm b}^2~=~e^{-3(\Nb - \Nd)}H_\dom^2$. The dilution factor during the EMD is then given by
\begin{equation} \label{eq:dilution_EMD}
    e^{-3(\Nb - \Nd)} = \frac{H_\text{b}^2}{H_\text{i}^2 \beta^4}~.
\end{equation}
The independence of $\rgh$ from $\beta$ becomes explicit when writing:
\begin{align} \label{eq:rho_gamma_H_indep}
    \rgh(\Nb) 
    &\approx \beta \rho_\tot (\Ni) e^{-3(\Nd - \Ni)} e^{-3(\Nb - \Nd)} \nonumber \\
    &\approx \beta H_\text{i}^2 \beta^3 \frac{H_\text{b}^2}{H_\text{i}^2 \beta^4} \approx H_\text{b}^2~.
\end{align}
Given that the relic density $\rho_\REM$ is fixed by $\rgh /  \mi$, it is also independent of the initial PBH abundance $\beta$. 

While requiring a generic, non-fine-tuned $\beta$ to induce an EMD era simply demands that the initial background redshifts faster than pressureless matter, the exact independence of $\rgh(\Nb)$ and $\rho_\REM(\Nb)$ from $\beta$ derived above relies on a stricter requirement: the early content of the Universe must redshift \textit{exactly} like radiation. This condition is naturally satisfied if the pre-EMD era is dominated by an oscillating scalar field acting as an inflaton after the end of inflation, provided it evolves in a quartic potential $V(\phi)\sim \phi^4$.

One can also calculate the reheating temperature after PBH evaporation.  At the $e$-fold time $\Nb$, a radiation density $\rho_{\rm rad}^{\rm H} = \rho_\gamma^{\rm H} + \rho_\nu^{\rm H}$ is injected into the Universe, which, after thermalization, corresponds to a bath at temperature $T$. This can be computed using the standard thermodynamic relation for a relativistic thermal gas:
\begin{equation} \label{eq:temperature_Nb}
    T (\rho_{\rm rad}^{\rm H},g_*)=  \left( \frac{30}{\pi^2} \frac{\rho_{\rm rad}^{\rm H}}{ g_*(T)} \right)^{1/4}~,
\end{equation}
where $g_*(T)$ is the effective number of relativistic degrees of freedom.  

\begin{figure*}
    \centering
    \includegraphics[width=1.\linewidth]{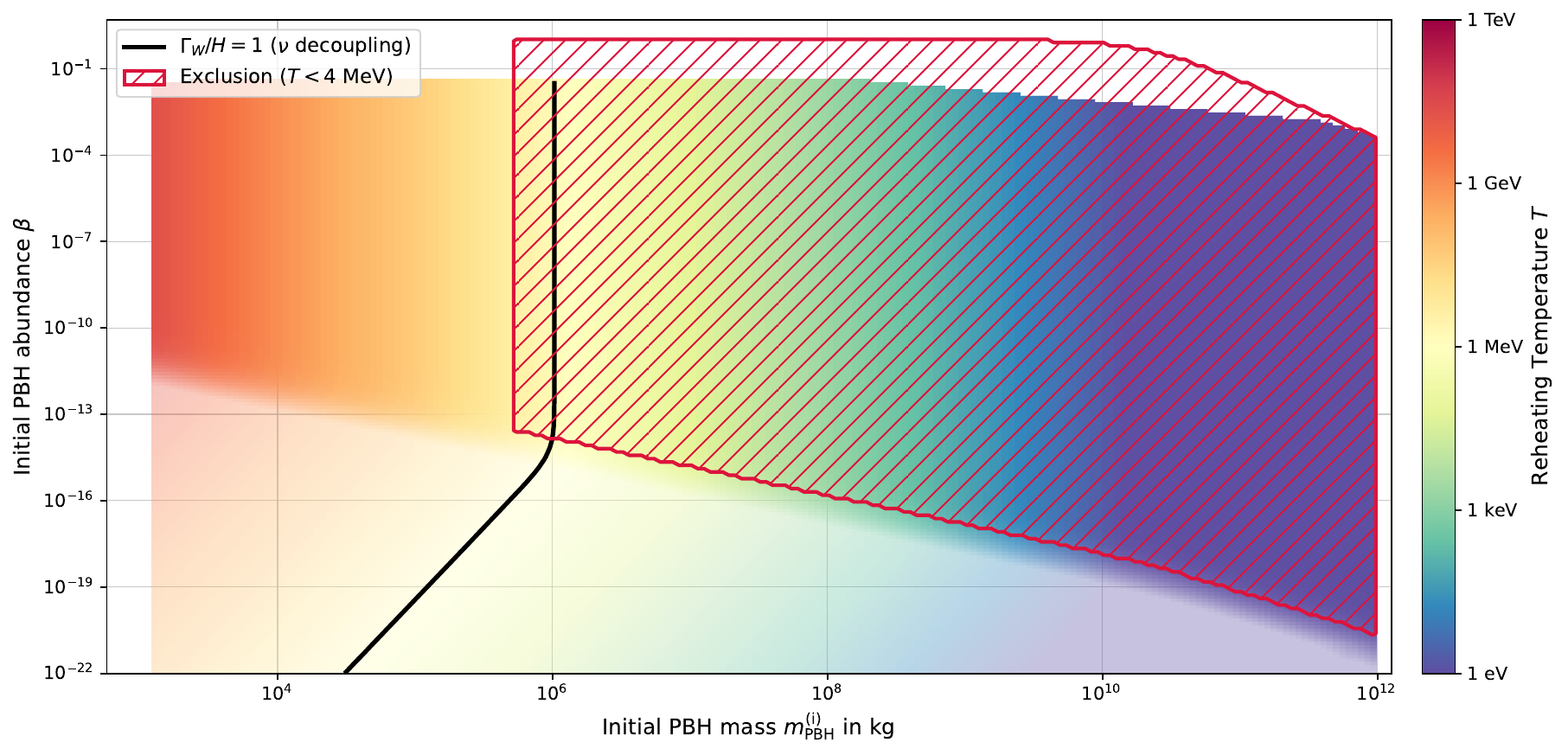}
    \caption{Reheating temperature $T_\text{reheat}$ of the thermal bath right after PBH evaporation (at $e$-fold time $\Nb$). The solid black curve ($\Gamma_{W}/H = 1$) indicates the neutrino decoupling temperature. The red hatched area highlights the excluded parameter space where $T_\text{reheat}(\Nb) <$ 4 MeV, conflicting with BBN constraints. In the lower region with faded colors, the initial fraction $\beta$ is insufficient to trigger an EMD era; consequently, the radiation budget remains dominated by the pre-existing primordial bath, no global reheating occurs at $\Nb$, and this thermal constraint becomes irrelevant.}
    \label{fig:reg_2_temperature}
\end{figure*}

Our results are shown in Fig.~\ref{fig:reg_2_temperature} displaying the reheating temperature as a function of $\mi$ and $\beta$.  As expected, it is independent of $\beta$, provided that there exists an EMD era.  Given that there is a minimum reheating temperature of $T_\text{reheat,min} \approx 4$ MeV \cite{Hannestad:2004px} from Big-Bang nucleosynthesis, one can exclude a wide range of PBH masses above $\mi\gtrsim 10^6$ kg.  This result is valid in general and not related to the production of Planckian relics. At the cosmological ``sweet spot'' $\mss$, the reheating temperature is orders of magnitude higher, around $T_{\rm reheat} \approx 100$ GeV. In this scenario, the PBHs safely reheat the Universe far above the BBN limit and even the QCD transition, retrieving the common thermal history of the Universe.  

\subsection{Thermalization Limits and Dark Radiation Constraints}\label{ssec:Thermalization Limits and Dark Radiation Constraints}

Computing the exact reheating temperature allows us to derive exclusion regions based on the thermalization efficiency of the injected particles. This efficiency depends heavily on the weak interaction rate $\Gamma_W$. If PBH evaporation occurs when $\Gamma_W / H < 1$, the injected neutrinos fail to couple to the thermal plasma. Rather than thermalizing, they contribute directly to dark radiation, sourcing an increase in the effective number of relativistic species, $\Delta \Neff$.

\begin{figure*}
    \centering
    \includegraphics[width=\linewidth]{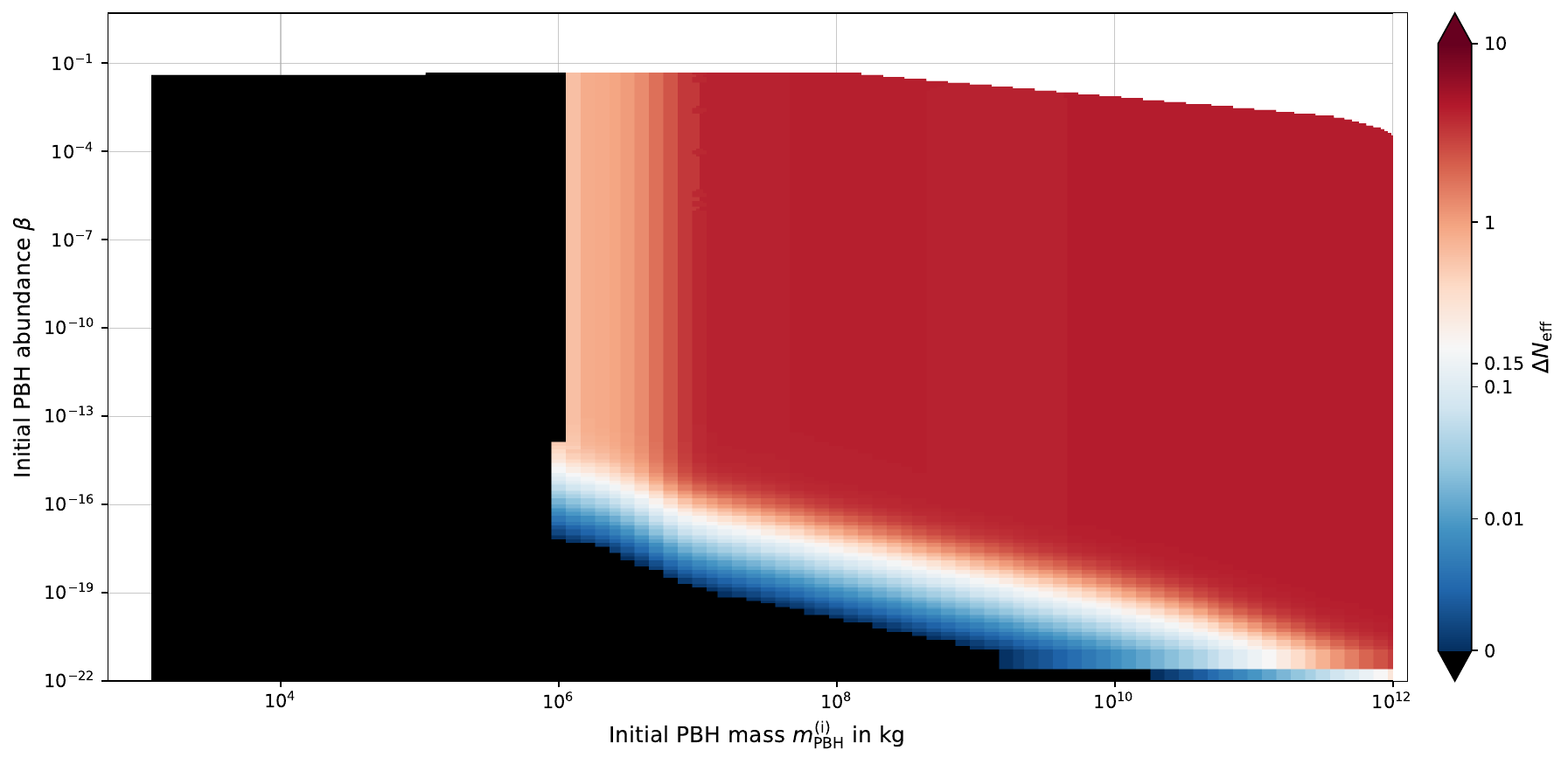}
    \caption{Variation of the effective number of relativistic degrees of freedom $\Delta \Neff$ from Hawking evaporation products. This excess is primarily sourced by neutrinos injected after the weak interaction decoupling temperature ($\Gamma_{W}/H < 1$), which fail to thermalize and act as dark radiation. The contribution from Hawking-emitted gravitons is included but remains highly subdominant.}
    \label{fig:results_reg2_Delta_Neff}
\end{figure*}

As illustrated in Fig.~\ref{fig:results_reg2_Delta_Neff}, the massive injection of non-thermalized neutrinos (and gravitons) significantly alters the standard radiation budget for heavy PBH masses, providing 
an upper bound on the initial abundance for masses above $10^6$ kg. The order of magnitude of the reheating temperature excluded by the $\Delta N_{\rm eff}$ bounds ($T \lesssim 1$ MeV) is similar to the threshold required to preserve primordial abundances during BBN ($T \gtrsim 4$ MeV) in Fig.~\ref{fig:reg_2_temperature}. Consequently, the exclusion zones derived from neutrino overproduction and standard BBN under-heating are essentially the same.

\section{Conclusion and perspectives}\label{sec:Conclusion and perspectives}

There exist multiple ways to form light PBHs in the early Universe~\cite{LISACosmologyWorkingGroup:2023njw}, e.g. from the growth of perturbations of an oscillating inflaton field, tachyonic or metric preheating, collapse of oscillons or topological defects, a peak in the primordial scalar power spectrum, a spectator field during inflation acting like a curvaton, a phase transition, etc. PBH formation is therefore no more speculative than other mechanisms involving physics beyond the Standard Model to explain DM and other cosmological or theoretical puzzles. 

The evaporation of light PBHs into Hawking radiation, which may have left stable Planckian remnants in LQG and other theoretical frameworks, provides a DM candidate as well as an alternative way to reheat the Universe after inflation.  These are the  motivations for this work, where we have systematically studied early-Universe cosmologies with light PBHs and Planckian remnants.  By confronting these scenarios with the present content of the Universe and with current and future constraints on primordial GW backgrounds, we have constrained the initial PBH abundance and the stability of remnants within the LQG framework, we have identified the required initial PBH mass and abundance to explain the DM, and we have excluded some scenarios as well.

Two regimes have been distinguished.  In the first one, corresponding to PBHs initially lighter than $10^3$ kg, it is possible that DM is made of Planckian remnants, on the condition that the density fraction collapsing in PBHs is tiny and fine-tuned.  In this regime, the Hawking evaporation products have only marginally contributed to the radiation content of the Universe and the cosmic evolution is standard.  It is hard to probe this regime observationally because it does not generate a detectable GW background. A relatively strong condition is obtained, as a function of the PBH mass, on the parameter $k$ describing the stability of remnants in Eq.~\ref{eq:tau tot}.  It ranges from $k >14$ for $\mi \sim 10^{-3}$ kg to $k>4$ for $\mi \sim 10^{3}$ kg.  Any firm theoretical prediction could therefore automatically rule out part of the PBH mass range for DM being constituted by Planckian remnants.  According to a recent development in this field~\cite{Bianchi:2026ytw}, the remnant lifetime may even scale as the exponential of the squared initial PBH mass and be much larger than the age of the Universe for any relevant mass, making Planckian remnants a viable DM candidate. 

In the second regime, corresponding to an initial PBH mass between $10^3$ kg and $10^{12}$ kg, we show that Planckian remnants cannot constitute all the DM.   The reason is that PBHs start to dominate the density of the Universe at some point and produce an EMD, at the end of which they evaporate and saturate the content of the Universe with Hawking radiation products.  Planckian relics are still present, but their maximal density today is well constrained, becoming more stringent as the PBH mass increases, down to $f_\REM \sim 10^{-24}$ for the upper bound of $10^{12}$ kg.   Moreover, in this regime, we have shown that most of the radiation content of the Universe originates from the Hawking evaporation, in a way that is  insensitive to the initial PBH abundance $\beta$.  This independence is also found for $f_\REM$. Furthermore, the PBH abundance $\beta$ is constrained by GW emission, boosted by the EMD era, induced either by scalar perturbations at the origin of PBH formation or by Poisson fluctuations in the distribution of PBHs.  It is possible to probe this regime with future GW observatories like LISA and the Einstein Telescope, and by the current and upcoming limits on the number of extra relativistic degrees of freedom $\Delta \Neff$.   In particular, depending on the PBH mass, $\beta$ values down to $10^{-5}$ are already ruled out by LVK and CMB limits on GWs from Poisson fluctuations.

Our analysis puts in evidence a particular case, referred to as a \textit{sweet spot}, at the transition between regime I and regime II, where all the DM can be made of Planckian remnants {\em and}\, PBH evaporation reheats the Universe. This occurs for a specific PBH mass around $10^3$ kg and we note that its exact value depends on the Barbero-Immirzi parameter $\gamma_{\text{\tiny LQG}}$, the only free parameter of LQG, fixing the remnant mass.  Any observational evidence for such a model would therefore provide a novel way to measure $\gamma_{\text{\tiny LQG}}$.  We also notice that the associated $\beta$ value does not need to be fine-tuned and can be between $10^{-11}$ and order one, and leads to observable GWs if $\beta \gtrsim 10^{-7}$.  This result is of particular interest in the context of PBH scenarios that typically require strong parameter fine-tunings.  

One possible way to produce PBHs with a mass of $10^3$ kg could be when the inflaton field oscillates at the bottom of the potential, leading to a growth of inflaton fluctuations that eventually collapse into PBHs.  If the equation of state in this phase is $w>0$ or if the inflaton decays into radiation before PBH domination, then we emphasize that the independence of $\beta$ is conserved. We plan to further study this particular scenario in a future work.   

This phenomenological analysis demonstrates that LQG offers a robust theoretical mechanism to rehabilitate light PBH remnants as DM candidates. By resolving the singularity and predicting a transition to stable Planckian remnants, LQG evades the strict constraints imposed by the semi-classical Hawking evaporation.

The cosmological scenarios studied here were restricted to the case of a monochromatic PBH mass and one could extend our analysis to more complex models with extended PBH mass abundances.  This may suppress the Poltergeist mechanism of GW production and increase the diversity of cosmological scenarios by altering the EMD phase. 
For masses above $10^6$ kg, we also pointed out that the reheating temperature defined by PBH evaporation is below 4 MeV and does not prevent the injection of non-thermalized high-energetic neutrinos, which additionally excludes a large part of the parameter space. 

Finally, an interesting perspective of this work would be to embed our scenario in models with a matter bounce replacing inflation, obtained in the context of LQG~\cite{Frion:2025cyd}. These could also lead to interesting alternative cosmologies and new ways to test the quantum nature of gravity.\\[1cm]

\subsection*{Acknowledgements}
AD and FV thanks all the ``Gravitation and Cosmology Group" at IEM-CSIC for discussions. AD is grateful to Mateo Pascual for insightful discussions and to Emmanuel Frion for his assistance with the code. The work of AD and FV was made possible through the support of the WOST, WithOut SpaceTime project (\href{https://withoutspacetime.org}{https://withoutspacetime.org}), supported by Grant ID\# 63683 from the John Templeton Foundation (JTF). The opinions expressed in this work are those of the authors and do not necessarily reflect the views of the John Templeton Foundation.\\

\appendix
\section{Computation of the \texorpdfstring{$\alpha$}{alpha}'s and the \texorpdfstring{$\beta$}{beta}'s}\label{app:Computation of the alpha parameters}
\subsection{Computation of the \texorpdfstring{$\alpha$}{alpha} parameters}
The $\alpha$ parameters are computed by requiring that the observed density of photons $\rho_\gamma^\obs$, baryonic matter $\rho_{\rm b}^\obs$ and DM $\rho^\obs_\text{\tiny DM}$ are consistent with the prediction of the model. The initial conditions are parametrized, for each background species $A$, by 
\begin{align}\label{eq:alpha_cond}
\rho_A (\Ni^-) \equiv \alpha_A e^{-3(1+w_A)\Ni} \rho_A^\obs~,
\end{align}
except for the neutrino density which is determined by the photon energy density, related by a factor $C_{\nu\gamma}$ assumed to be $\frac{7}{8} \Neff (4/11)^{4/3}$ with $\Neff = 2.99$. $\Ni^-$ is the $e$-fold time just before PBH formation. One obtains the $\alpha$ parameters by solving Eq.~\ref{eq:alpha_cond} together with the equations governing each phase described in section~\ref{sec:PBH cosmologies in LQG}.

For the phase \PI, the $\alpha$ parameters only depend on $\Ni$, the initial abundance $\beta$ and the observed densities:
\begin{align}
    \alpha_b^\PI        &= \frac{1}{1 - \beta}~,\\
    \alpha_c^\PI        &=  1 - \frac{\beta}{1 - \beta}\frac{\rho_b^\text{obs}+ \rho_\gamma^\text{obs} (1+ C_{\nu\gamma}) e^{- \Ni}}{\rho_\DM^\text{obs}}~,\\
    \alpha_\gamma^\PI   &= \frac{1}{1 - \beta}~.
\end{align}

In phase \PII, $\alpha$ also depends on $\Nb$, the $e$-fold time corresponding to the ``decay" of PBHs, on $\epsilon$, the fraction of energy going into the remnants, and on $\epsilon_A$, $A\in \{e,\,p,\,\gamma\,,\nu\,,h\}$, the fraction of energy density from Hawking radiation going into each species $A$.
\begin{align}
    \alpha_b^\PII &=\frac{1}{1-\beta}\label{eq:alphaP2b}~,
\end{align}
\begin{widetext}
\begin{align}
    \alpha_c^\PII&=
    \frac{1}{1-\beta}
    +\frac{\beta \epsilon}{(\beta -1)\rho_\DM^\obs}
    \frac{
        e^{\Ni} \left( \rho_b^\obs + \rho_\DM^\obs \right)
        + \left( 1+C_{\nu\gamma}\right) \rho_\gamma^\obs
    }{
        e^{\Ni} \left( 1- \beta [1-\epsilon] \right) 
        + e^{\Nb} \beta (1+C_{\nu\gamma}) (1-\epsilon) \epsilon_\gamma
    }
    \label{eq:alphaP2c}~,\\
    \alpha_\gamma^\PII&=
    \frac{e^{\Ni}}{(\beta-1)\,\rho_\gamma^\obs}
        \frac{%
            \rho_\gamma^\obs \left( 1 - \beta[1-\epsilon]\right)
            -e^{\Nb} \beta(1-\epsilon)\epsilon_\gamma\bigl(\rho_b^\obs+\rho_\DM^\obs\bigr)
        }{%
            e^{\Ni}\!\bigl(-1+\beta[1-\epsilon]\bigr)
            -(1+C_{\nu\gamma})e^{\Nb}\beta(1-\epsilon)\epsilon_\gamma
        }~.
\end{align}
\end{widetext}
In the last phase $\texttt{P3}$, the parameters additionally depend on $\Nr$, the $e$-fold time corresponding to the ``decay" of the remnants.
\begin{widetext}
\begin{minipage}{0.48\textwidth}
\begin{equation}
    \alpha_b^\PIII = \frac{1}{1-\beta}~,
\end{equation}
\end{minipage}\hfill
\begin{minipage}{0.48\textwidth}
\begin{equation}
    \alpha_c^\PIII = \frac{1}{1-\beta}~,
\end{equation}
\end{minipage}
    \begin{align}
    \alpha_\gamma^\PIII = 
    \frac{e^{\Ni}}{(1-\beta)\rho_\gamma^\obs}
    \frac{
        (1-\beta) \rho_\gamma^\obs
        -(\rho_b^\obs + \rho_\DM^\obs) \beta  \left( 
            e^{N_r} \epsilon 
        +e^{\Nb} [1-\epsilon] \epsilon_\gamma 
        \right)
    }{
        e^{\Ni} (1-\beta)
        +(1+\C)\beta \left(
            e^{N_r} \epsilon
            +e^{\Nb} [1-\epsilon] \epsilon_\gamma
        \right)
    }~.
    \end{align}
\end{widetext}
One gets the maximal allowed values for $\beta$ in each saturation regime and in each phase by solving Eq.~\ref{eq:beta_max_cdt}. For $\alpha_c^\PI$, $\alpha_\gamma^\PII$ and $\alpha_c^\PII$, one gets:
\begin{widetext}
\begin{minipage}{0.48\textwidth}
\begin{equation}
    \beta_c^\PI = \frac{\rho_\DM^\obs}{\rho_b^\obs + \rho_\DM^\obs + \left(1+C_{\nu\mu}\right) e^{\Ni} \rho_\gamma^\obs}~,
\end{equation}
\end{minipage}\hfill
\begin{minipage}{0.48\textwidth}
\begin{equation}
    \label{eq:beta_gamma}
    \beta_\gamma^\PII = \frac{\rho_\gamma^\obs}{\left(1-\epsilon\right)\left( e^{\Nb}\epsilon_\gamma \left[ \rho_b^\obs + \rho_\DM^\obs \right] + \rho_\gamma^\obs \right)}~,
\end{equation}
\end{minipage}
\begin{align}\label{eq:beta_c}
    \beta_c^\PII = \frac{\rho_\DM^\obs}{
    \rho_\DM^\obs + \epsilon \rho_b^\obs
    -\left(1+C_{\nu\gamma}\right)e^{\Nb-\Ni} \left(1-\epsilon\right)\epsilon_\gamma \rho_\DM^\obs
    +e^{-\Ni}\left(1+C_{\nu\gamma}\right)\epsilon \rho_\gamma^\obs
    }~.
\end{align}
\end{widetext}
\newpage
\section{Numerical approach to the boundary value problem}\label{app:Numerical and temperature}

A difficulty arises in the photon saturation regime when the solver must determine a vanishingly small initial radiation parameter $\alpha_\gamma^\PII$ by subtracting the Hawking contribution from the present-day photon density. While this regime is relevant whenever an EMD era is induced, the numerical precision limit is severely hit for initial PBH masses $\mi \gtrsim 10^8$ kg at high initial abundances $\beta$, as shown in Fig.~\ref{fig:results_LQG}. In this specific mass range, the Hawking radiation component scales as $e^{\Nb-\Ni} \sim 10^{15}-10^{21}$ during the PBH-dominated era, with the duration $\Nb - \Ni$ growing with the PBH mass.

By definition of the initial conditions (Eq.~\ref{eq:alpha_cond}), the parameter $\alpha_\gamma^\PII$ represents the exact fraction of the present-day photon density originating from the pre-existing plasma. It is therefore physically identical to the deficit parameter introduced in Sec.~\ref{subsec:Genericity of beta in regime II} ($\delta_\gamma \equiv \alpha_\gamma^\PII$). Due to the architecture of the code, $\alpha_\gamma^\PII$ is never exactly zero but is minimized to an arbitrarily small quantity $\delta$. When the numerical solver reaches its precision limit, it effectively enforces an artificial lower bound $\delta_\gamma \ge \delta$, which prevents the simulation from reaching the pure photon-saturation attractor ($\delta_\gamma \to 0$).

To isolate and treat this numerical artifact, we modify the theoretical prediction for the extremized abundance $\beta_\gamma$ by solving the equation $\alpha_\gamma^\PII = \delta$. This yields a modified prediction for the photon-saturation threshold, $\beta_\gamma^{(\delta)}$, replacing Eq.~\ref{eq:beta_gamma}:
\begin{equation}\label{eq:beta_gamma_delta}
    \beta_\gamma^{(\delta)} = \frac{\mathcal{A} + \sqrt{\mathcal{C} + \mathcal{D}^2}}{\mathcal{B}}~,
\end{equation}
\begin{widetext}
where the terms are defined as:
\begin{align}
    \mathcal{A} &= -(1 + C_{\nu\gamma}) e^{\Nb} \delta (1 - \epsilon) \epsilon_\gamma \rho_\gamma^\obs - e^{\Ni} \big(e^{\Nb} [1 - \epsilon] \epsilon_\gamma [\rho_b^\obs + \rho_\DM^\obs] + [1 - 2 \delta] \rho_\gamma^\obs - [1 - \delta] \epsilon \rho_\gamma^\obs\big)~, \\
    \mathcal{B} &= 2 \delta (1 - \epsilon) (e^{\Ni} - [1 + C_{\nu\gamma}] e^{\Nb} \epsilon_\gamma) \rho_\gamma^\obs~, \\
    \mathcal{C} &= 4 e^{\Ni} (1 - \delta) \delta (1 - \epsilon) (e^{\Ni}- [1 + C_{\nu\gamma}] e^{\Nb} \epsilon_\gamma){(\rho_\gamma^\obs)}^2~, \\
    \mathcal{D} &= e^{\Ni} \big(-[1 - \epsilon] + \delta [2  - \epsilon ]\big) \rho_\gamma^\obs - e^{\Nb + \Ni} (1 - \epsilon) \epsilon_\gamma (\rho_b^\obs + \rho_\DM^\obs) - (1 + C_{\nu\gamma}) e^{\Nb}\delta (1 - \epsilon) \epsilon_\gamma \rho_\gamma^\obs~.
\end{align}
\end{widetext}

Despite this numerical limitation, the physical results remain robust. We validate the numerical constraints by comparing the analytical predictions of $\beta_\gamma$ (Eq.~\ref{eq:beta_gamma}) and the residue-corrected $\beta_\gamma^{(\delta)}$ (Eq.~\ref{eq:beta_gamma_delta}) against the simulation output $\beta(\mi)$.

\begin{figure*}
    \centering
    \includegraphics[width=\linewidth]{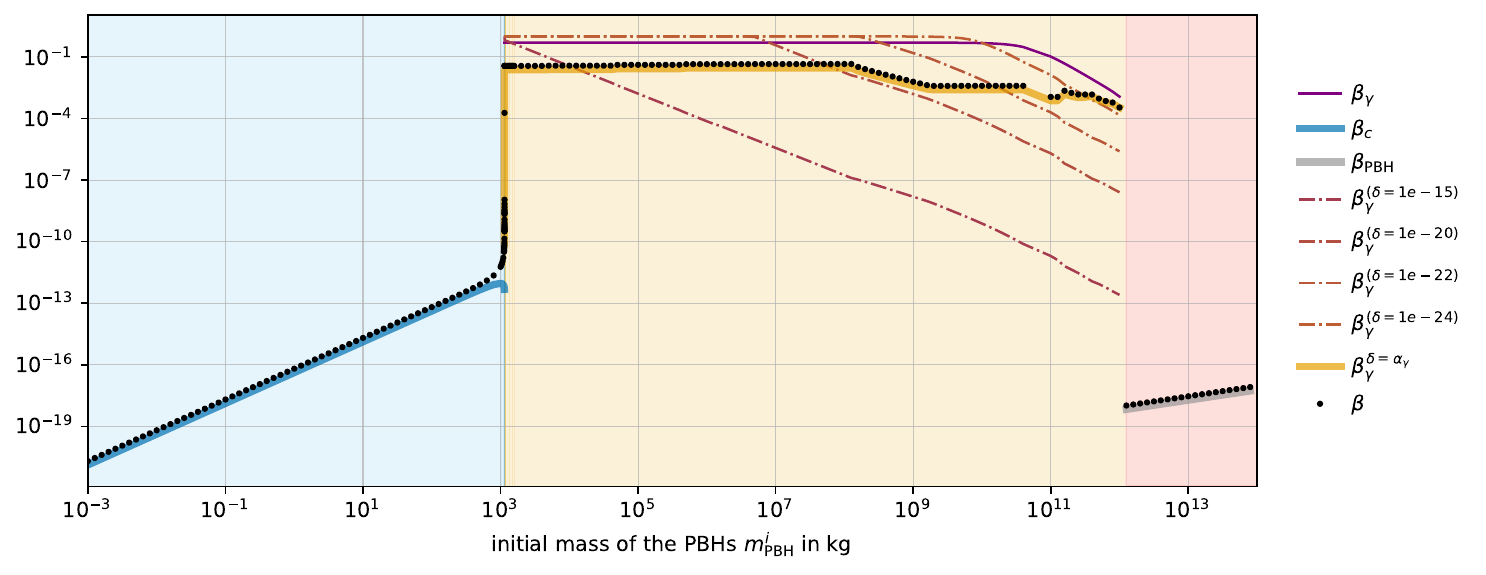}
    \caption{
    Simulation results for the maximum initial PBH abundance $\beta$ is compared with the semi-analytical solutions derived in Apps.~\ref{app:Computation of the alpha parameters} and~\ref{app:Numerical and temperature}. The thick blue curve $\beta_c$ (Eq.~\ref{eq:beta_c}) corresponds to the DM saturation regime, while the purple curve $\beta_\gamma$ (Eq.~\ref{eq:beta_gamma}) represents the photon saturation regime. The dashed lines $\beta_\gamma^{(\delta)}$ (Eq.~\ref{eq:beta_gamma_delta}) illustrate the analytical prediction when incorporating a numerical residue $\delta$, showing that the noise observed in the high-mass regime ($\mi \gtrsim 10^5$ kg) is a numerical artifact caused by the vanishingly small value of $\alpha_\gamma^\PII$ required in the photon-saturated era. When evaluated on the specific residue obtained numerically, $\beta_\gamma^{(\delta = \alpha_\gamma^\PII)}$ exactly recovers the simulation results.
    }
    \label{fig:result_comparison}
\end{figure*}

As shown in Fig.~\ref{fig:result_comparison}, we compare the theoretical values $\beta_c$ and $\beta_\gamma$ with the simulation points. For completeness, we include $\beta_\pbh$, computed analogously to $\beta_c$ for phase P1. A discrepancy appears in the photon-saturated regime between the pure theoretical limit $\beta_\gamma$ and the simulation $\beta(\mi)$. However, when we plot $\beta_\gamma^{(\delta)}$ evaluated using the exact numerical residue $\delta = \alpha^\PII_\gamma(\mi)$ obtained iteratively for each PBH mass, the discrepancy vanishes, perfectly matching the simulation results.

Finally, it is important to emphasize that the numerical framework remains indispensable despite the existence of these analytical checks. The analytical expressions for $\beta_c$ explicitly depend on the formation and evaporation epochs, $\Ni$ and $\Nb$. These are not free parameters but dynamic variables determined by the integrated history of the Hubble rate, $H(N)$. The simulation is therefore required to capture the relative dilutions and the energy transfers between distinct fluids, ensuring that the early-Universe dynamics consistently converge toward the present-day observational bounds.

\bibliography{ref_clean}
\end{document}